\documentclass[journal,twocolumn]{IEEEtran}
\usepackage{epsfig,makeidx,color}
\usepackage{amsmath,amssymb,bbm}
\usepackage{cite,graphicx,lipsum}
\usepackage{enumerate}
\usepackage[switch,pagewise]{lineno}
\usepackage{hyperref}
\hypersetup{
        colorlinks = true,
        citecolor=blue,
}
\def\defh{\mbox{\footnotesize $ \begin{array}{c} H_0 \cr > \cr < \cr H_1 \end{array} $}}
\def\defhr{\mbox{\footnotesize $ \begin{array}{c} H_1 \cr > \cr < \cr H_0 \end{array} $}}

\def\cK{{\cal K}}

\def\cQ{{\cal Q}}

\def\rH{{\rm H}}
\def\rT{{\rm T}}

\def\uP{{\mathbb P}}

\def\uC{{\mathbb C}}
\def\uE{{\mathbb E}}

\newtheorem{mylemma}{\bf Lemma} 
\newtheorem{myexample}{\it Example} 

\def\be{ \begin{equation} }
\def\ee{ \end{equation} }
\def\bea{ \begin{eqnarray} }
\def\eea{ \end{eqnarray} }

\def\bx{{\bf x}}

\def\bc{{\bf c}}
\def\bd{{\bf d}}
\def\bb{{\bf b}}

\def\bg{{\bf g}}

\def\bs{{\bf s}}
\def\ba{{\bf a}}

\def\bm{{\bf m}}
\def\bn{{\bf n}}
\def\bh{{\bf h}}

\def\bp{{\bf p}}

\def\bee{{\bf e}}

\def\bv{{\bf v}}

\def\bE{{\bf E}}

\def\bI{{\bf I}}

\def\bN{{\bf N}}

\def\bQ{{\bf Q}}
\def\bR{{\bf R}}

\def\bY{{\bf Y}}
\def\bZ{{\bf Z}}

\def\b0{{\bf 0}}
\def\bPhi{{\bf \Phi}}

\def\cA{{\cal A}}
\def\cC{{\cal C}}

\def\cQ{{\cal Q}}

\def\cN{{\cal N}}

\def\sSINR{{\sf SINR}}

\ifCLASSOPTIONonecolumn
  \newcommand{\figwidth}{0.50\columnwidth}
  
\else
  \newcommand{\figwidth}{0.85\columnwidth}
  
\fi

\begin{document}

\title{An Approach to Preamble Collision Reduction 
in Grant-Free Random Access with Massive MIMO}

\author{Jinho Choi\\
\thanks{The author is with
the School of Information Technology,
Deakin University, Geelong, VIC 3220, Australia
(e-mail: jinho.choi@deakin.edu.au).
This research was supported
by the Australian Government through the Australian Research
Council's Discovery Projects funding scheme (DP200100391).}}


\maketitle
\begin{abstract}
In this paper, we study grant-free random
access with massive multiple input multiple output (MIMO)
systems. We first show that the performance
of massive MIMO based grant-free random access 
is mainly decided by the probability of preamble collision.
The implication of this is that although
the number of antennas can be arbitrarily large,
the (average) number of successful packet transmissions 
can be limited by the number of preambles.
Then, we propose an approach to preamble collision reduction 
without increasing the number of preambles
using the notion of superpositioned preambles (S-preambles).
Since the channel estimation for conjugate beamforming
can be carried out when unused S-preambles are known,
a simple approach 
is derived to detect unused S-preambles and its performance is analyzed
for a special case (superpositions with 2 preambles).
Based on analysis and simulation results,
it is confirmed that the proposed approach with
S-preambles can increase the success probability with 
the same spectral efficiency as the conventional approach.
\end{abstract}


\ifCLASSOPTIONonecolumn
\baselineskip 28pt
\fi

\section{Introduction}

Machine-type communication (MTC) 
has been extensively studied as it plays a crucial role
in providing the connectivity for devices in 
5th generation (5G) cellular systems and
the Internet of Things (IoT)
\cite{Bockelmann16} \cite{Dawy17}.
In MTC, although there are a large number of devices to be connected,
in general, they have short data packets with sporadic activity.
Thus, random access is employed for MTC thanks to low signaling overhead
\cite{3GPP_MTC} \cite{3GPP_NBIoT}
\cite{Chang15} \cite{Choi16}.

To support massive connectivity for
a large number of IoT devices and sensors,
the notion of massive multiple input multiple output (MIMO)
\cite{Marzetta10} can be considered
with a base station (BS) equipped with a large number of antenna
elements.
Based on \cite{Bjornson18}, it seems that massive MIMO is a solution to
massive MTC, as the capacity becomes unbounded in the presence
of pilot contamination, which may allow to support a very large
number of devices in each cell.
In particular, various random access schemes with massive MIMO
are studied for massive MTC in \cite{deC17}  \cite{Senel17} \cite{Liu18}
\cite{Ding19_IoT},
where grant-free random access 
is considered to take advantage of 
high spatial resolution or selectivity in massive MIMO.
Note that grant-free random access is also called
two-step random access or one-shot approaches \cite{Bockelmann18}
\cite{Choi17IoT} \cite{Choi20b}.

In \cite{deC17} \cite{Liu18} \cite{Ding19_IoT},
the random access schemes with massive MIMO have two phases
for grant-free random access. 
In the first phase, an active device
sends a preamble that is randomly chosen from a predetermined
preamble pool (i.e., a finite set of sequences)
so that the BS is able to estimate the channel state information
(CSI) of the device. In the second phase,
the active device transmits its data packet and
the BS performs beamforming
with estimated CSI to decode data packets
from multiple active devices.
The resulting scheme is grant-free random access
as no handshaking for request-grant is employed
(i.e., the BS does not send any feedback signal to active
devices for granting access or reserving channels).
Provided that each active device's channel 
is different from others (due to 
high spatial resolution or selectivity in massive MIMO),
the BS is able to form multiple beams to decode
data packets from multiple active devices,
which makes 
grant-free random access with massive MIMO suitable for massive
connectivity.

In grant-free 
random access with massive MIMO,
there are various issues including
device activity detection \cite{Liu18}
and preamble design \cite{Jiang19}.
Among those,
as discussed in \cite{Jiang19},
preamble design plays a key role
in providing a good performance in terms of the probability of successful
decoding or success probability.
In this paper, since the success probability mainly depends
on the probability of
preamble collision,
without increasing the size of preamble pool,
we propose an approach to preamble collision reduction,
which results in the decrease of
the probability of preamble collision or the increase
of success probability.
In particular,
using superposition of a limited number of orthogonal sequences,
a large number of preambles,
which are referred to as superpositioned preambles (S-preambles),
are generated to lower the the probability of preamble collision.
Furthermore, thanks to the fact that each S-preamble
is a sum of orthogonal sequences, it is possible to
estimate the channel vectors without interference from other preambles,
which leads to a better performance
than that of conventional non-orthogonal preambles
(e.g., Zadoff-Chu (ZC) or Gaussian random sequences) \cite{Ding20b}. 
With a special case of superpositioned preambles 
(i.e., superpositions with two preambles),
we derive a low-complexity preamble detection method and
analyze the performance in the paper.

Note that in \cite{Jiang19}, 
transmitting multiple \emph{consecutive} preambles
is considered to increase the success probability,
which unfortunately
leads to the increase of the length of preamble transmission phase
and a poor spectral efficiency.
On the other hand, in the proposed approach, the 
length of the preamble transmission phase
is the same as that of preamble (as in conventional approaches,
e.g., \cite{Ding19_IoT}).
In other words, with the same spectral efficiency
as the conventional grant-free random access with massive MIMO,
a higher success probability can be achieved.

It is also possible to decrease
the probability of preamble collision by increasing the number of preambles.
With a fixed length of preambles (or a fixed
system bandwidth), it is possible to 
have a large number of preambles if preambles are not orthogonal.
In this case,
the preamble detection can be seen as a sparse
signal recovery problem in the context of compressive sensing 
(CS) \cite{Donoho06} \cite{Candes06}, and various CS algorithms
can be applied to the preamble detection (or user activity detection)
as in \cite{Senel17} \cite{Liu18} \cite{Seo19} \cite{Choi19c}.
In general, the complexity of CS algorithms is high
(usually proportional to the number of preambles) for
joint activity detection. 
On the other hand,
in this paper, since a set of orthogonal sequences is used
to increase the number of preambles,
simple low-complexity
correlators can be used for the preamble detection,
which means that the complexity is low as it is dependent on the length
of preambles. Consequently,
S-preambles 
allow us not only to use the low-complexity channel estimation 
(as the case of orthogonal preambles), 
but also to lower the probability of collision 
(as the case of non-orthogonal preambles), 
which results in a better performance.

The rest of the paper is organized
as follows. In Section~\ref{S:SM},
we present the system model for
grant-free random access with massive MIMO,
and show that the performance is mainly limited
by the size of preamble pool.
Then, we propose an approach to preamble collision
reduction without increasing the size of preamble pool
(or the length of preamble transmission phase)
in Section~\ref{S:EA} and 
derive a hypothesis testing approach to detect
unused preambles that is necessary for the channel
estimation in Section~\ref{S:Hyp}.
In Section~\ref{S:Sim}, we present simulation results.
The paper is concluded in Section~\ref{S:Con} with some remarks.

\subsubsection*{Notation}
Matrices and vectors are denoted by upper- and lower-case
boldface letters, respectively.
The superscripts $\rT$ and $\rH$
denote the transpose and complex conjugate, respectively.
The Kronecker product is denoted by $\otimes$.
$\uE[\cdot]$
and ${\rm Var}(\cdot)$
denote the statistical expectation and variance, respectively.
$\cC \cN(\ba, \bR)$
represents the distribution of
circularly symmetric complex Gaussian (CSCG)
random vectors with mean vector $\ba$ and
covariance matrix $\bR$.
The Q-function is given by
$\cQ(x) = \int_x^\infty \frac{1}{\sqrt{2 \pi} } e^{- \frac{t^2}{2} } dt$.

\section{System Model}	\label{S:SM}

Suppose that a grant-free random access system consists of
a BS and multiple devices.
It is assumed that the BS is equipped with
$M$ antenna elements and each device has a single antenna.

\subsection{Grant-free Random Access}

\begin{figure}[thb]
\begin{center}
\includegraphics[width=\figwidth]{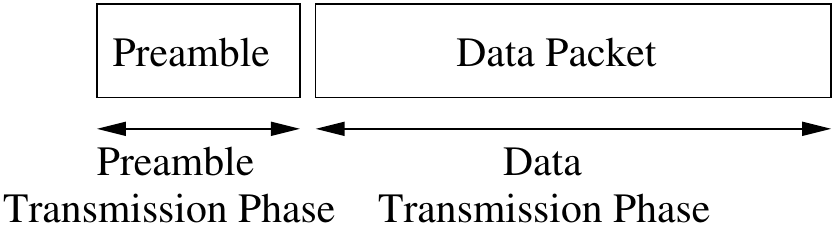}
\end{center}
\caption{Two phases
(i.e., preamble transmission and data transmission
phases) for a grant-free random access scheme.}
        \label{Fig:two_phase}
\end{figure}

In grant-free or 2-step random access,
there are two phases \cite{Ding19_IoT}
\cite{Jiang19}. The first phase is the preamble
transmission phase and the following phase is the data
transmission phase as illustrated in Fig.~\ref{Fig:two_phase}.
In the preamble transmission phase,
each active device with data is to send
a randomly selected preamble from the following preamble pool:
$\cC = \{\bc_1, \ldots, \bc_L\}$,
where $\bc_l \in \uC^{N \times 1}$ represents the $l$th preamble. 
Throughout the paper, 
it is assumed that the $\bc_l$'s are orthonormal sequences
of length $N$ and $|[\bc_l]_n| = \frac{1}{\sqrt{N}}$ for all $l$
and $n$.
Thus, we assume that $L = N$.
After sending a preamble, an active device
sends its data packet in
the data transmission phase as shown in  
Fig.~\ref{Fig:two_phase}.
Throughout the paper, it is assumed that the 
lengths of data packets of devices
are the same. In addition, all the devices
are synchronized (to this end, the BS needs periodically broadcast
a beacon signal).

Let $\bh_k \in \uC^{M \times 1}$ represent the channel vector 
from active device $k$ to the BS,
which remains unchanged for the 
transmission slot consisting of consecutive
preamble and data transmission phases.
Then, in the preamble transmission phase,
the BS receives the following signal in the space
and time domain:
\be
\bY = \sum_{k=1}^K \bh_k
\sqrt{P_k} \bc_{l(k)}^\rT + \bN \in \uC^{M \times N},
\ee
where $K$ represents the number of active devices,
$l(k)$ denotes the index of the preamble chosen
by active device $k$, 
$P_k$ is the transmit power of active device $k$,
and 
$[\bN]_{m,n} \sim \cC \cN(0, N_0)$ is the background noise
at the $m$th antenna and the $n$th preamble symbol duration.

Each active device transmits its 
data packet of length $D$ during the data transmission phase.
The corresponding received signal at the BS becomes
\be
\bZ =  \sum_{k=1}^K \bh_k
\sqrt{P_k} \bb_k^\rT + \tilde \bN \in \uC^{M \times D},
\ee
where $\bb_k$ represents the data packet from active device $k$
and 
$[\tilde \bN]_{m,d} \sim \cC \cN(0, N_0)$ is the background noise
at the $m$th antenna and the $n$th data symbol duration.
Throughout the paper, we assume that
$\uE[\bb_k] = \b0$ and $\uE[\bb_k \bb_k^\rH] = \bI$
(i.e., the symbol energy is normalized).

The BS uses a bank of 
$L$ correlators to detect
transmitted preambles and estimate the channel vectors
of the devices that associate with the detected preambles.
The output of the correlator with $\bc_l$ is given by
\begin{align}
\bg_l 
& = \bY \bc_l 
= \sum_{k=1}^K \bh_k \sqrt{P_k} \delta_{l(k),l} + \bN \bc_l \cr
& = \sum_{k \in \cK_l} \bh_k \sqrt{P_k} + \bN \bc_l, 
	\label{EQ:correlator}
\end{align}
where $\delta_{l,l^\prime}$
is the Kronecker delta (i.e.,
$\delta_{l,l^\prime} = 1$ if $l = l^\prime$, and 0 otherwise)
and $\cK_l$ represents the index set of the active
devices that choose preamble $l$.
If active device $k$ is the only 
device that chooses preamble $l$ (i.e., $l(k) = l$ and
$\cK_l  = \{k\}$),
thanks to the orthogonality of preambles,
it can be shown that
\be
\bg_l = \bh_k \sqrt{P_k} + \bn_l,
	\label{EQ:gn}
\ee
where $\bn_l = \bN \bc_l \sim \cC \cN \left(\b0, 
N_0 \bI \right)$.
To decode the data packet from 
active device $k$, conjugate
beamforming is applied and the output
of the beamformer becomes
\begin{align}
\bx_l & = \bg_l^\rH \bZ \cr
& = \bg_l^\rH \bh_k \sqrt{P_k} \bb_k +
\sum_{k^\prime \ne k} \bg_l^\rH \bh_{k^\prime} 
\sqrt{P_{k^\prime}}\bb_{k^\prime}^\rT + 
\bg_l^\rH \tilde \bN.
	\label{EQ:bd_l}
\end{align}
If $|\bg_l^\rH \bh_k|^2$ is sufficiently larger
than $|\bg_l^\rH \bh_{k^\prime}|^2$, 
a high signal-to-interference-plus-noise ratio (SINR)
can be achieved for successful decoding.

Throughout this paper, as in \cite{Marzetta10},
let $\bh_k = \sqrt{\ell_k} \bv_k$,
where $\ell_k$ represents
the large-scale fading coefficient that depends
on the distance between the $k$th active device
and the BS and $\bv_k$ 
stands for the small-scale
fading vector. 
For tractable analysis, 
we consider the following assumption \cite{Bjornson16}.
\begin{itemize}
\item[{\bf A)}] 
Throughout the paper,
$P_k$ is decided to be inversely proportional to 
$\ell_k$ via power control so that
\be
\bh_k \sqrt{P_k} = \bv_k \sqrt{P_{\rm rx}},
	\label{EQ:A}
\ee
where $P_{\rm rx}$ represents the (average) receive signal power
and $\bv_k \sim \cC \cN(\b0, \bI)$ is independent for all
$k$ (i.e., Rayleigh fading is assumed for small-scale fading).
\end{itemize}
In addition, for convenience, let
$\gamma  = \frac{P_{\rm rx}}{N_0}$ be the 
signal-to-noise ratio (SNR).

\subsection{Size of Preamble Pool and Preamble Collision} 

Suppose that all $K$ active devices randomly choose preambles
from $\cC$. The probability that one active device is not collided with
the other active devices
or one active device is collision-free
is given by
$ \uP_{\rm cf} 
= \left(1 - \frac{1}{L} \right)^{K-1} \le e^{-\frac{K-1}{L}}$.
For active device $k$,
the probability
that its signal is successfully decoded 
(i.e., the SINR is higher than
a decoding threshold without preamble collision)
is given by
\be
\eta_k = \uP_{\rm cf} \Pr(\sSINR_k \ge \Omega),
	\label{EQ:eta_k}
\ee
where $\Omega > 0$ is the SINR threshold for successful decoding.
For convenience, in short, $\eta_k$ is referred to
as the success probability.
In \cite{Ding19_IoT}, closed-form expressions for
$\Pr(\sSINR_k \ge \Omega)$ are derived with various 
beamforming approaches.
In general, if $M$ 
is sufficiently large, for a finite $\Omega$,
$\Pr(\sSINR_k \ge \Omega)$ approaches 1 (as the SINR grows
linearly with $M$ on average).
Thus, we can see that the performance limiting factor
(for each active device) is 
$\uP_{\rm cf}$ that depends on
the size of preamble pool, $L$.
Thus, for orthogonal preambles, to increase $L$,
the length of preambles needs to increase,
which unfortunately results in a poor spectral efficiency
(e.g., as in \cite{Jiang19}).

\section{An Effective Approach to Preamble Collision Reduction}
\label{S:EA}

In this section, we propose an approach
to preamble collision reduction
without increasing the size of preamble pool, $L$,
which can effectively improve
the success probability of each active device.

\subsection{Transmission of Multiple Preambles per Device}

While each active device is to choose only one
preamble to transmit in the preamble transmission phase
in the conventional approach,
we propose to choose $B~ (\ge 1)$ 
multiple preambles by each active device.
An active device is to send a superposition of $B$ different preambles,
which is called a S-preamble.
Then, the number of possible selections becomes
$Q = \binom{L}{B}$.
Let $\cA_q$ denote the index set of 
the $q$th selection, $q = 1, \ldots, Q$. 
For example, with $L = 4$ and $B = 2$, 
we can have
\begin{align}
& \cA_1 = \{1,2\}, 
\cA_2 = \{1,3\}, 
\cA_3 = \{1,4\}, 
\cA_4 = \{2,3\}, \cr
& \quad \cA_5 = \{2,4\}, 
\ \mbox{and} \ 
\cA_6 = \{3,4\}.
	\label{EQ:426}
\end{align}
Then, the $q$th S-preamble can be defined as
\be
\bs_q = \sum_{l \in \cA_q} \bc_l.
\ee
In the proposed approach,
the received 
signal in the preamble transmission phase 
is given by
\be
\bY = \sum_{k=1}^K \bh_k \bs_{q(k)}^\rT + \bN \in \uC^{M \times N},
	\label{EQ:25}
\ee
where $q(k)$ denotes the S-preamble index 
chosen by active device $k$.
For convenience, let $\cQ$ be the index set of the S-preambles
chosen by $K$ active devices.

In summary, the proposed approach is a grant-free
or 2-step approach where an active device transmits
a sum of $B$ orthogonal preambles 
in the preamble transmission phase. Thus, it has the following steps:
\begin{itemize}
\item[S1)] In the first step, an active device
chooses one of $Q$ S-preambles uniformly at random and transmits it,
which is a sum of $B$ orthogonal preambles,
followed by a data packet transmitted 
in the data transmission
phase as shown in Fig.~\ref{Fig:two_phase}. 
\item[S2)] In the second step,
the BS is to detect transmitted preambles and estimate
the channel vectors of the active devices
associated with transmitted preambles. 
In Subsection~\ref{SS:CE}, we will discuss a low-complexity 
channel estimation method by taking advantage of S-preambles.
With the estimated
channel vectors, the BS decodes the data packets
and sends feedback signals (of decoding outcomes)
to active devices.
\end{itemize}

It is clear that the probability that
one active device is not collided with
the other active devices in terms of S-preambles
is given by
\begin{align}
\uP_{\rm cf} 
= \left(1 - \frac{1}{Q} \right)^{K-1}.
	\label{EQ:Pnc2}
\end{align}
For example, suppose that $L = 64$ and $K = 10$.
Then, if $B = 1$ (i.e., the conventional approach),
the probability that one active device is collision-free 
is $\left(1 -\frac{1}{L}\right)^{K-1} \approx 0.867$.
When S-preambles are used with $B = 2$ in the proposed approach,
this probability becomes $\left(1 -\frac{1}{Q}\right)^{K-1} \approx 0.995$.

\begin{figure}[thb]
\begin{center}
\includegraphics[width=\figwidth]{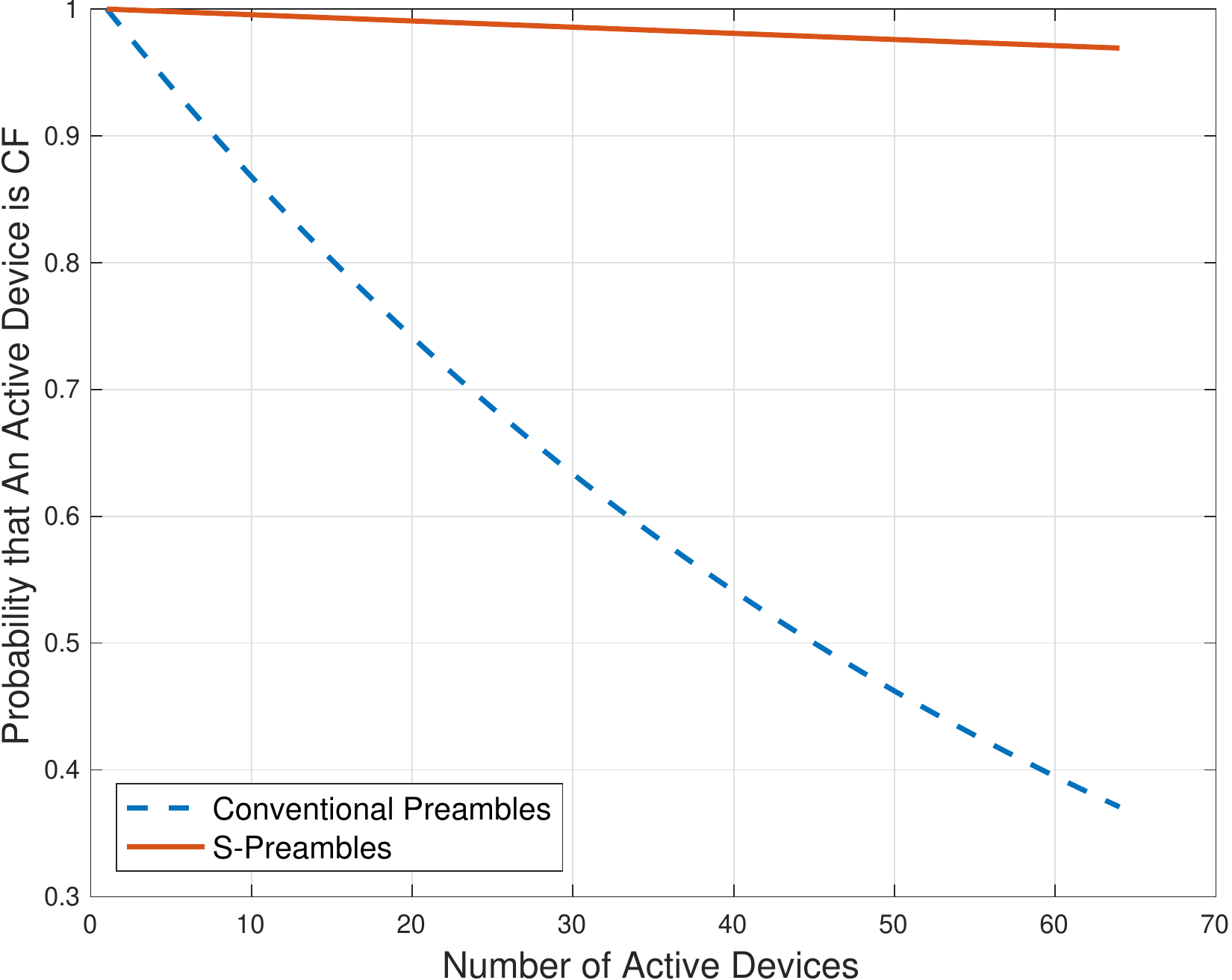}
\end{center}
\caption{Probability that one active device is collision-free
in terms of S-preambles
for different numbers of active devices when
$L = 64$ and $B = 2$.}
        \label{Fig:pnc}
\end{figure}

In Fig.~\ref{Fig:pnc},
we show 
the probability that one active device is not 
collided with the other active devices
for different numbers of active devices when
$L = 64$ and $B = 2$.
Clearly, using S-preambles,
the probability becomes significantly high,
which results in the increase of
the success probability as shown in \eqref{EQ:eta_k}.
Note that in \eqref{EQ:eta_k},
$\Pr(\sSINR_k \ge \Omega)$ is 
not related to the type of preambles.

Since $Q$ increases with $B~
\left(\le \lfloor \frac{L}{2} \rfloor\right)$, we can see that
the probability decreases with $B$.
Therefore, it is possible to reduce preamble collisions
and improve the performance.

\subsection{Channel Estimation with Known Transmitted S-Preambles}
\label{SS:CE}

In the rest of the paper, for simplicity,
we mainly focus on the case that $B = 2$
and discuss the channel estimation
with known transmitted S-preambles.

For convenience, suppose that each S-preamble is associated with one device
(without any collision). In addition,
the device associated with the $q$th S-preamble is 
referred to as the $q$th virtual active device.
The associated virtual channel vector is denoted by $\ba_q$. 
If the $q$th S-preamble is transmitted, the $q$th virtual
active device is one of active devices. On the other hand,
if the $q$th S-preamble is not transmitted, the $q$th virtual
active device is not an active device. 
With $Q$ virtual active devices, the received
signal in the preamble transmission phase can be rewritten as
\be
\bY = \sum_{q=1}^Q \sqrt{P_{\rm rx}} \ba_q \bs_q^\rT + \bN.
	\label{EQ:bY2}
\ee

In order to show how S-preambles work together with
virtual channel vectors, 
it might be useful to consider
a simple example with $L = 4$.
Then, as mentioned earlier, 
there are $Q = \binom{4}{2} = 6$ possible choices
as in \eqref{EQ:426}.
In this case, without background noise and receive power term,
from \eqref{EQ:bY2}, we have
\begin{align*}
\bg_l = \bY \bc_l 
= 
\left(\sum_{q} \ba_q \left(\sum_{l^\prime \in \cA_q} 
\bc_{l^\prime}\right)^\rT
\right) \bc_l.
\end{align*}
For example, we have $\bg_1 = \ba_1 + \ba_2 + \ba_3$,
$\bg_2 = \ba_1 + \ba_4 + \ba_5$, and so on. 
Thus, it can be shown that
\be
\bg = 
\left[
\begin{array}{c}
\bg_1 \cr
\bg_2 \cr
\bg_3 \cr
\bg_4 \cr
\end{array}
\right] 
= (\bPhi \otimes \bI)
\left[ \ba_1^\rT \
\ba_2^\rT \
\ba_3^\rT \
\ba_4^\rT \
\ba_5^\rT \
\ba_6^\rT \right]^\rT,
\ee
where
$$
\bPhi = 
\left[
\begin{array}{llllll}
1 & 1 & 1 & 0 & 0 & 0 \cr
1 & 0 & 0 & 1 & 1 & 0 \cr
0 & 1 & 0 & 1 & 0 & 1 \cr
0 & 0 & 1 & 0 & 1 & 1 \cr
\end{array}
\right].
$$
In general, the size of $\bPhi$ is $L \times Q$.
If a bipartite graph 
with $L$ left vertices and $Q$ right vertices
is considered, 
$\bPhi$ can be seen as a transfer or biadjacency matrix
\cite{Asratian98}.
Furthermore, it can be shown that
the number of 1's of each column is 2 
and that of each row is $L-1$. Thus, $\bPhi$
is the transfer matrix of
a bipartite graph with $L$ left vertices and $Q$ right vertices 
that is $(L-1)$-regular on the left and $2$-regular on the right.

Note that if $B$ is greater than 2,
$\bPhi$ becomes the transfer matrix of
a bipartite graph with $L$ left vertices and $Q = 
\binom{L}{B}$ right vertices 
that is $\binom{L-1}{B-1}$-regular on the left and $B$-regular on the right.
Thus, for $B \ge 3$, the resulting bipartite graph
becomes an expander graph. 
Since we only focus on the case of $B = 2$ in this paper,
a generalization with $B \ge 3$ is to be studied in the future
as a further research topic.

In \eqref{EQ:bY2}, since the $\ba_q$'s are virtual,
if no active device chooses the $q$th S-preamble,
we have $\ba_q = \b0$ and it is not necessary to estimate 
the corresponding channel.
In other words,
we only need to estimate the channels associated
with the transmitted (or chosen) S-preambles.
For convenience, let 
$\bar K$ be the number of the virtual channels 
associated with the transmitted S-preambles.
If there is no preamble collision, $\bar K$ becomes $K$.
Suppose that $\bar K \le L$ and denote by $\bar \bPhi$
the submatrix of $\bPhi$ obtained by taking
the columns corresponding to the transmitted S-preambles.
Then, for example, using the least squares (LS) approach \cite{Scharf91},
the $\bar K$ channel vectors can be estimated as follows:
\be
\hat \ba = (\bar \bPhi^\dagger \otimes \bI) \bg,
	\label{EQ:LS}
\ee
where $\bar \bPhi^\dagger$
represents the Moore–Penrose pseudoinverse
of $\bar \bPhi$ and $\hat \ba$ is a vector of
the estimates of $\bar K$ virtual channel vectors.

Clearly, in order to estimate the channel vectors as above,
it is important to detect \emph{unused} S-preambles.
We discuss an approach based
on the outputs of correlators as follows.

With $B = 2$,
recall
that $\cA_q = \{q_1, q_2\}$ is the set of the two indices of the preambles
used in the $q$th S-preamble, i.e., $q_1$ and $q_2$,
where $q_b \in \{1,\ldots, L\}$.
For the $q$th S-preamble, the BS can use
the following correlators' outputs:
\begin{align}
\bg_{q,1} = \bY \bc_{q_1} \ \mbox{and} \ 
\bg_{q,2} = \bY \bc_{q_2}.
	\label{EQ:29}
\end{align}

\begin{myexample}	\label{E:1}
With $L = 4$,
suppose that there are 3 active devices, i.e., $K = 3$,
that choose S-preambles $1$, $3$, and $4$, i.e.,
$\cQ = \{1,3,4\}$.
In this case, without background noise and receive power term,
it can be shown that
\begin{align*}
q=1: \ \{\bg_{1,1}, \bg_{1,2}\} & = \{\bv_1+\bv_2, \bv_1+ \bv_3\} \cr
q=2: \ \{\bg_{2,1}, \bg_{2,2}\} & = \{\bv_1+\bv_2, \bv_3\} \cr
q=3: \ \{\bg_{3,1}, \bg_{3,2}\} & = \{\bv_1+\bv_2, \bv_2\} \cr
q=4: \ \{\bg_{4,1}, \bg_{4,2}\} & = \{\bv_1+\bv_3, \bv_3\} \cr
q=5: \ \{\bg_{5,1}, \bg_{5,2}\} & = \{\bv_1+\bv_3, \bv_2\} \cr
q=6: \ \{\bg_{6,1}, \bg_{6,2}\} & = \{\bv_3, \bv_2\}.
\end{align*}
\end{myexample}

Since active device 1 chooses 
the first S-preamble,
in both $\bg_{1,1}$ and $\bg_{1,2}$,
$\bv_1$ can be found.
For the pairs of $\{\bg_{q,1},\bg_{q,2}\}$ corresponding to the
S-preambles chosen by the other active devices,
we have the same observation.
On the other hand, 
the pairs of $\{\bg_{q,1},\bg_{q,2}\}$ corresponding to the
S-preambles not chosen by any active devices,
there are no common channel vectors
in both $\bg_{q,1}$ and $\bg_{q,2}$.
Based on the above observation, we have the following result.

\begin{mylemma}	\label{L:3}
From \eqref{EQ:25} and \eqref{EQ:29},
it can be shown that
\begin{align}
\bg_{q,1} & = \sum_{k=1}^K \bv_k \sqrt{P_{\rm rx}}
\delta_{\cA_{q (k)}(1), q_1} + \bN \bc_{q_1} \cr
\bg_{q,2} & = \sum_{k=1}^K \bv_k \sqrt{P_{\rm rx}}
\delta_{\cA_{q (k)}(2), q_2} + \bN \bc_{q_2},
\end{align}
where $\cA_q (b)$ represents the $b$th element of $\cA_q$.
Then, if the $q$th S-preamble is chosen
by $K_q$ active devices, 
under the assumption of {\bf A)},
w.p. 1, we have
\be
\lim_{M \to \infty} \frac{\bg_{q,1}^\rH \bg_{q,2}}{M} \to 
P_{\rm rx} K_q.
	\label{EQ:L3}
\ee
\end{mylemma}
\begin{IEEEproof}
For a given S-preamble, say S-preamble $q$,
since $q_1$ and $q_2$ are different,
based on the strong law of large numbers,
w.p. 1, it follows
$$
\frac{(\bN \bc_{q_1})^\rH \bN \bc_{q_2} }{M} \to 0,  
$$
as $M \to \infty$.
Similarly, under the assumption of {\bf A)},
$$
\frac{ \left(\sum_{k=1}^K \bv_k \sqrt{P_{\rm rx}}
\delta_{\cA_{q (k)}(b), q_b}\right)^\rH \bN \bc_{q_{b^\prime}}}{M} \to 0,
$$
as $M \to \infty$,
for $b \ne b^\prime$.
Since there are $K_q$ common channel vectors
in both $\sum_{k=1}^K \bv_k \sqrt{P_{\rm rx}}
\delta_{\cA_{q (k)}(b), q_b}$, $b = 1,2$,
we can show that \eqref{EQ:L3} holds.
\end{IEEEproof}

Consequently, from \eqref{EQ:L3}, the correlation between
$\bg_{q,1}$ and $\bg_{q,2}$ can be used 
to decide whether
or not the $q$th S-preamble is chosen by any active device.
If the unused S-preambles are detected, the corresponding virtual
channels are assumed to be zero vectors. 
As in Example~\ref{E:1},
we can see that the correlation 
coefficients between $\bg_{q,1}$ and $\bg_{q,2}$
corresponding to $q \in \{2,5,6\}$
would approach 0. Thus, there are non-zero virtual 
channel vectors for $q \in \{1,3,4\}$. 
Then, we have
$$
\bar \bPhi =
\left[
\begin{array}{llllll}
1 &  1 & 0  \cr
1 &  0 & 1  \cr
0 &  0 & 1  \cr
0 &  1 & 0  \cr
\end{array}
\right]
$$
in \eqref{EQ:LS} to estimate non-zero virtual channel
vectors, $\hat \ba$.
Without any S-preamble collision,
the number of columns of $\bar \bPhi$ is equivalent
to the number of active devices.
Thus, it is possible that
the channel vectors of the
active devices can be estimated 
when $K \le L$ without S-preamble collision.
However, the condition that
$K \le L$ without S-preamble collision is not a sufficient
condition for the channel estimation, 
because the rank of $\bar \bPhi$
may not be the same as the number of columns (when $K \le L$).
For example, when $\cQ = \{1,2,5,6\}$ with $K = 4$ and $L = 4$.
the rank of $\bar \bPhi$ is 3, not 4,
which means that the channel estimation for $K = 4$
active devices is not possible.

\begin{figure}[thb]
\begin{center}
\includegraphics[width=\figwidth]{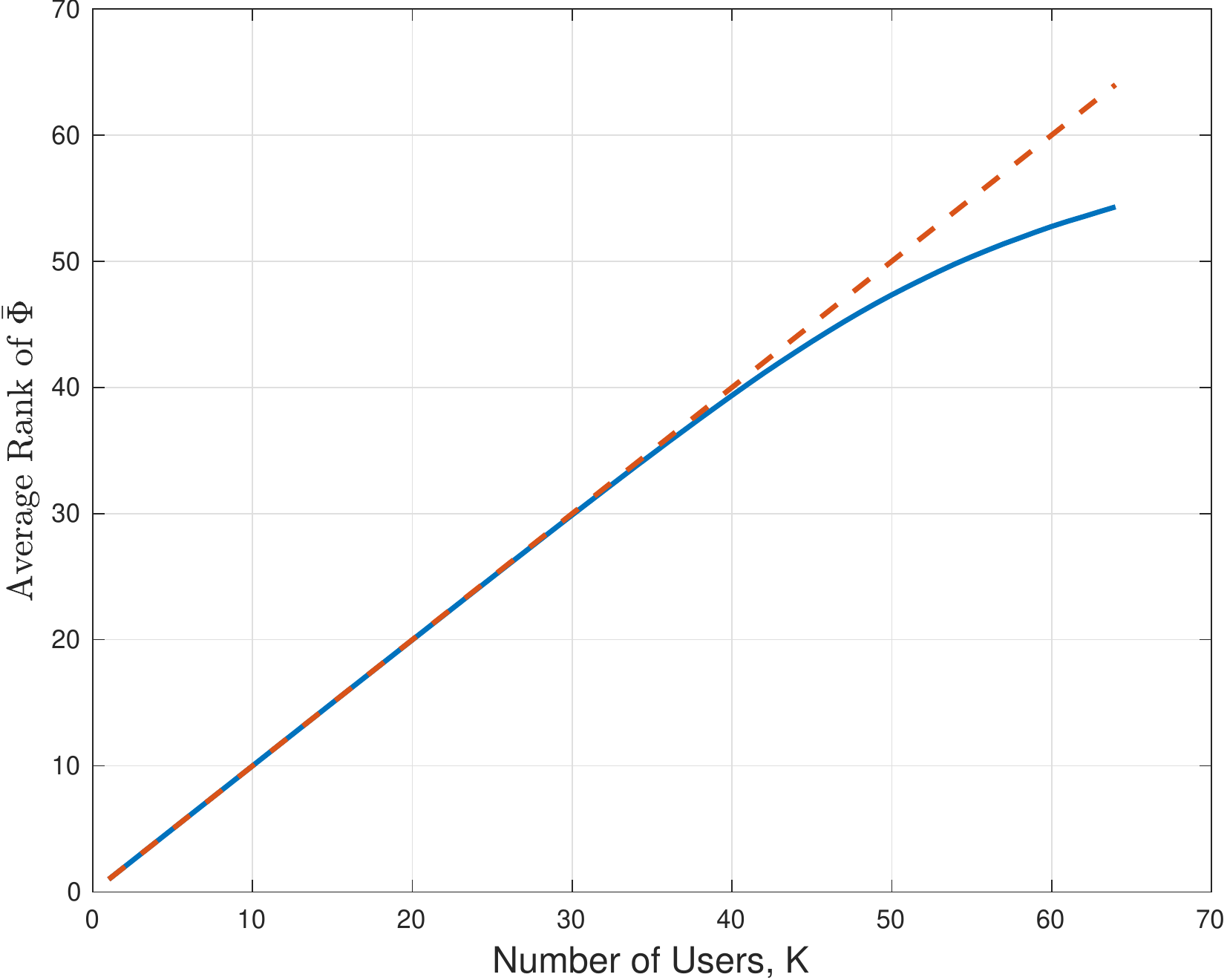}
\end{center}
\caption{The average rank of $\bar \bPhi$ as a function
of $K$ when when $K = L = 64$, which is shown by
the solid line (the dashed line represents
the number of the columns of $\bar \bPhi$, i.e., $K$).}
        \label{Fig:t_mat}
\end{figure}

In Fig.~\ref{Fig:t_mat}, 
the (empirical) average rank of $\bar \bPhi$,
$\uE[{\rm rank} (\bar \bPhi)]$, is shown
as a function of $K$ when $L = 64$
when $K$ (out of $Q$) columns 
of $\bPhi$ are randomly selected for $\bar \bPhi$.
It is shown that the difference between the rank of 
$\bar \bPhi$ and $K$ increases as $K$ approaches $L$.
Thus, although the use of S-preamble can mitigate
the preamble collision, 
due to rank deficient $\bar \bPhi$ for a $K$ close
to $L$, some channels cannot be 
estimated. In Section~\ref{S:Sim}, we 
present simulation results that show the
impact of rank deficient $\bar \bPhi$ on the 
success probability.

There are a few remarks.

\begin{itemize}
\item To increase the number of preambles, 
non-orthogonal preambles can be used.
For example, ZC or Gaussian random sequences 
can be used for preambles \cite{Ding20b}.
As shown in \cite{Ding20b},
although the number of preambles can be large
(to lower the probability of collision),
its performance (in terms of success probability)
can be worse than that of orthogonal preambles
due to the correlation between preambles that results in 
degraded channel estimates and SINR.

\item It is noteworthy that the channel estimation
in \eqref{EQ:LS} is based on the outputs of simple correlators
in \eqref{EQ:correlator}.
As a result, the complexity is mainly proportional to
the number of multiplications to carry out $\bY \bc_l$ for all $l$,
which is $L^2 M$.
In addition, to find the unused S-preambles,
we need to find the correlation between the $\bg_l$'s, 
which requires a complexity order of $O(L^2 M)$ with $B = 2$.
Thus, the proposed approach with S-preambles has a low computational 
complexity (i.e., $O(L^2 M)$).
Note that as mentioned earlier,
CS algorithms can be used to detect
non-orthogonal preambles (including S-preambles)
as in \cite{Senel17} \cite{Liu18} \cite{Seo19} \cite{Choi19c}.
Since the complexity of CS algorithms depends on 
the length of preambles as well as the number of preambles, the 
resulting complexity can be high (e.g., a
variational inference algorithm used in \cite{Choi19c}
requires a complexity of $O(Q M L^2)$ per iteration.
Thus, with $B = 2$, we have $Q = O(L^2)$, 
which means that the complexity order
becomes $O(M L^4)$ per iteration.
Clearly, S-preambles allow us not only to use the low-complexity
channel estimation studied in this section
for grant-free random access with massive MIMO,
but also to lower the probability of collision,
which results in a better performance.

\end{itemize}

\section{Hypothesis Testing 
for the Detection of Unused Preambles}	\label{S:Hyp}

Since Lemma~\ref{L:3} is valid when $M \to \infty$,
with a finite $M$ (but sufficiently large),
a hypothesis testing approach \cite{Kay98}
is required for the detection
of unused S-preambles (or transmitted S-preambles).
To this end,
let $D = \frac{\bg_{q,1}^\rH \bg_{q,2}}{\sqrt{M}}$.
The distributions of the test statistic $D$
can be found from \cite{Mallik11}.
For a large $M$, alternatively, we can use
the central limit theorem (CLT).
For convenience, we only consider the $q$th S-preamble.
In addition, we assume that
the number of the active devices choosing 
preamble $q_b$, $b \in \{1,2\}$, associated with
S-preamble $q$ is known,
which is denoted by $\kappa_{q_b}$.
For example, consider Example~\ref{E:1}
with $A_q = \{q_1, q_2\} = \{1,2\}$ when $q =1$. 
In this case, $\kappa_{q_1} = 2$ and $\kappa_{q_2} = 2$.

Let $H_0$ represent
the hypothesis that S-preamble $q$ is unused
and $H_1$ the alternative hypothesis.
If $\kappa_{q_b} = 0$ for any $b \in \{1,2\}$,
it is clear that S-preamble $q$ is not chosen. 
Thus, we only focus on the case that
$\kappa_{q_b} > 0$ for $b \in \{1,2\}$.
In addition, for $H_1$, we only consider the case that $K_q = 1$
(as the probability that $K_q \ge 2$ might be sufficiently low).

Under $H_0$,  based on the CLT, it is readily shown that
$\uE[D] = 0$,
${\rm Var}(D) = (P_{\rm rx} \kappa_{q_1} +N_0)(P_{\rm rx} \kappa_{q_2} +N_0)$,
and $\uE[D^2] = 0$. 
For convenience, let $\rho_b = P_{\rm rx} \kappa_{q_b} + N_0$.
Then, under $H_0$, 
$D \sim \cC \cN \left(0, \rho_1 \rho_2 \right)$.
The distribution of $D$ under $H_1$ can be found as follows.
\begin{mylemma}
Using the CLT, under $H_1$, with $K_q= 1$, it can be shown that
\be
D \sim \cC \cN \left(\sqrt{M} P_{\rm rx}, 
\rho_1 \rho_2, P_{\rm rx}^2 \right),
\ee
where $X \sim \cC \cN(\uE[X], {\rm Var}(X),
\uE[X^2] - \uE[X]^2)$ represents a complex Gaussian
random variable  \cite{Picinbono96}.
Note that while $D$ under $H_0$
is CSCG, $D$ under $H_1$ is not CSCG (as $\uE[D^2] - \uE[D]^2 \ne 0$).
\end{mylemma}
\begin{IEEEproof}
With $K_q = 1$, suppose that the $k$th active device chooses
the $q$th S-preamble. Then,
the $m$th elements of $\bg_{q,b}$, $b \in \{1,2\}$, are given by
\begin{align}
[\bg_{q,1}]_m & = \sqrt{P_{\rm rx}} [\bv_k]_m  + \nu_{1,m} + n_{1,m} \cr
[\bg_{q,2}]_m & = \sqrt{P_{\rm rx}} [\bv_k]_m  + \nu_{2,m} + n_{2,m},
\end{align}
where $n_{b,m}$ represents the $m$th element of
$\bN \bc_{q_b}$ and
$\nu_{b,m} \sim \cC \cN(0, 
P_{\rm rx} (\kappa_{q_b}- 1))$.
As a result, $\nu_{b,m} + n_{b,m} \sim \cC \cN(0, 
P_{\rm rx} (\kappa_{q_b}- 1) + N_0)$.
For convenience, let $I_b = P_{\rm rx} (\kappa_{q_b}- 1) + N_0$.

Since each of $\bv_k \sim \cC \cN(\b0, \bI)$ is iid,
and $n_{1,m}$ and $n_{2,m}$ are independent
zero-mean CSCG random variables,
we have
\begin{align}
\uE[D] 
= \uE \left[
\frac{1}{\sqrt{M}}
\sum_m ([\bg_{q,1}]_m )^* ([\bg_{q,2}]_m )
\right] 
= \sqrt{M} P_{\rm rx}.
\end{align}
It can also be shown that
\begin{align}
\uE[|D|^2] 
& = \frac{1}{M}
\uE \left[\bigl|\sum_m ([\bg_{q,1}]_m )^* ([\bg_{q,2}]_m ) \bigl|^2 \right]\cr
& = 3 P_{\rm rx}^2  +
(M-1) P_{\rm rx}^2 
+ (I_1+ I_2) P_{\rm rx} + I_1 I_2, \ \ 
\end{align}
by using the fact that $\uE[|X|^4] = 2 \sigma^4$ 
for $X \sim \cC \cN(0,\sigma^2)$ .
As a result, the variance of $D$ is given by
\begin{align}
{\rm Var}(D) 
= P_{\rm rx}^2 + (I_1+ I_2) P_{\rm rx} + I_1 I_2  = \rho_1 \rho_2.
	\label{EQ:VD1}
\end{align}
It can also be readily shown that
\be
\uE[(D-\uE[D])^2] = P_{\rm rx}^2,
	\label{EQ:VD2}
\ee
which completes the proof.
\end{IEEEproof}

Let $\bd = [D \ D^*]^\rT$.
In addition, let
$\bR_0 = {\rm diag}(\rho_1 \rho_2, \rho_1 \rho_2)$ and
\be
\bR_1 = \left[
\begin{array}{cc}
\rho_1 \rho_2 & P_{\rm rx}^2 \cr
P_{\rm rx}^2  & \rho_1 \rho_2 \cr
\end{array}
\right].
\ee
Then, according to  \cite{Bos95} \cite{Picinbono96},
it can be shown
that
\be
f(\bd\,|\, H_i) = \frac{1}{\pi (\det \bR_i)^{-\frac{1}{2}} }
e^{
- \frac{1}{2} (\bd - \bm_i)^\rH \bR_i^{-1} (\bd - \bm_i)},
\ee
for $i \in \{0,1\}$,
where $\bm_0 = [0 \ 0]^\rT$ and
$\bm_1 = [ \sqrt{M} P_{\rm rx} \ \sqrt{M} P_{\rm rx} ]^\rT$.
As a result, 
using the log-likelihood ratio (LLR),
the hypothesis testing becomes
\be
T(\bd) = 
(\bd - \bm_1)^\rH \bR_1^{-1} (\bd - \bm_1)
-\bd^\rH \bR_0^{-1} \bd 
\defh \lambda,
	\label{EQ:U_lam}
\ee
where $\lambda > 0$ is a decision 
threshold.
While the LLR test can be 
carried out as in \eqref{EQ:U_lam},
it is not straightforward to find a decision boundary 
and error probabilities as $\bR_0 \ne \bR_1$ \cite{Duda00}.
However, thanks to the symmetric property of $\bR_1$,
the following result becomes useful
to simplify the hypothesis testing.

\begin{mylemma}
The test statistic in \eqref{EQ:U_lam}
can be expressed as
\be
T(\bd) = 
\frac{|y_1 - u_1|^2}{\sigma_1^2} -
\frac{|y_1|^2}{\rho_1 \rho_2} 
+ \left( \frac{1}{\sigma_2^2} - \frac{1}{\rho_1 \rho_2} \right) |y_2|^2, \ 
	\label{EQ:L5}
\ee
where $y_1 = \sqrt{2} \Re(D)$,
$y_2 = \sqrt{2} \Im(D)$, 
$u_1 = \sqrt{2 M} P_{\rm rx}$,
$\sigma_1^2 = \rho_1 \rho_2+ P_{\rm rx}^2$,
and $\sigma_2^2 = \rho_1 \rho_2 - P_{\rm rx}^2$.
\end{mylemma}
\begin{IEEEproof}
It can be shown that
the eigenvalue decomposition of $\bR_1$ becomes
$\bR_1 = \bE~{\rm diag}(\sigma_1^2, \sigma_2^2) \bE^\rH$,
where 
$$
\bE = [\bee_1 \ \bee_2] = 
\frac{1}{\sqrt{2}} \left[ \begin{array}{cc} 1 & 1 \cr
1 & -1 \cr \end{array} \right].
$$
Here, $\bee_i$ represents the eigenvector of $\bR_1$
corresponding to the eigenvalue $\sigma_i^2$.
Then, it is readily shown that
$y_i = \bee_i^\rH \bd$, $i = 1,2$. In addition,
$\bee_1^\rH \bm_1 = u_1$ and $\bee_2^\rH \bm_1 = 0$. 
As a result, it can be shown that
\begin{align}
& (\bd - \bm_1)^\rH \bR_1^{-1} (\bd - \bm_1) \cr
& = [(y_1 - u_1) \ y_2]^* 
{\rm diag}\left(\frac{1}{\sigma_1^2}, 
\frac{1}{\sigma_2^2} \right) 
[(y_1 - u_1) \ y_2]^\rT,
	\label{EQ:L5_1}
\end{align}
while 
\be
\bd^\rH \bR_0^{-1} \bd = 
\frac{1}{\rho_1 \rho_2} \bd^\rH \bE \bE^\rH \bd = 
\frac{1}{\rho_1 \rho_2} (|y_1|^2 + |y_2|^2).
	\label{EQ:L5_2}
\ee
Then, from \eqref{EQ:L5_1} and \eqref{EQ:L5_2}, we have \eqref{EQ:L5},
which completes the proof.
\end{IEEEproof}

From \eqref{EQ:L5},
it can be seen that $y_1$ plays a more significant role
in the LLR test than $y_2$.
Thus, the following simplified test statistic can be found:
\be
\bar T(y_1) = 
\frac{|y_1 - u_1|^2}{\sigma_1^2} - \frac{|y_1|^2}{\rho_1 \rho_2}.
	\label{EQ:bT}
\ee

In fact, as shown below, $\bar T(y_1)$ is the LLR
with $y_1$.

\begin{mylemma}	\label{L:6}
Based on the CLT, under $H_1$ with $K_q = 1$, we have
\be
y_1 = \sqrt{2} \Re(D) \sim \cN( u_1, \sigma_1^2).
\ee
In addition, $y_1 \sim \cN(0, 
\sigma_0^2 )$ under $H_0$,
where $\sigma_0^2 = \rho_1 \rho_2$. 
\end{mylemma}
\begin{IEEEproof}
Since $D$ is a CSCG random variable under $H_0$,
it can be readily shown that the variance of $\sqrt{2}\Re(D)$
is $2 \frac{{\rm Var}(D)}{2} = {\rm Var} (D) = \rho_1 \rho_2$.
Under $H_1$, from \eqref{EQ:VD1} and \eqref{EQ:VD2}, it can be shown that
\begin{align*}
{\rm Var} (D) & = {\rm Var} (\Re(D)) + {\rm Var}(\Im (D)) = 
\rho_1 \rho_2 \cr
\uE[(D - \uE[D])^2] & = {\rm Var} (\Re(D)) - {\rm Var}(\Im (D))  \cr
& \quad + j 2 \uE[ \Re(D - \uE[D]) \Im (D- \uE[D]) ]
= P_{\rm rx}^2,
\end{align*}
which implies that $\uE[ \Re(D - \uE[D]) \Im (D- \uE[D]) ] = 0$,
and ${\rm Var}(\Re(D)) = \frac{\rho_1 \rho_2 + P_{\rm rx}^2}{2}$.
From this, we can show that the variance of $y_1$ is $\sigma_1^2$.
Furthermore, the mean of $y_1$ is $\sqrt{2} \uE[\Re(D)] = \sqrt{2} \uE[D] = 
u_1$. These complete the proof.
\end{IEEEproof}

From Lemma~\ref{L:6}, with the test statistic in \eqref{EQ:bT},
from \cite{Kay98},
the LLR test can be simplified as follows:
\be
y_1 \defhr \tau,
	\label{EQ:y1t}
\ee
where $\tau$ is a decision threshold.
Then, the probabilities of false alarm (FA)
and missed detection (MD), denoted by $\uP_{\rm FA}$
and $\uP_{\rm MD}$, respectively,
in the S-preamble  detection associated with \eqref{EQ:y1t}
can be found as
\begin{align}
\uP_{\rm FA} & 
= \cQ \left( \frac{\tau}{\sqrt{\sigma_0^2}}\right) 
= \cQ \left( \frac{\tau}{\sqrt{\rho_1 \rho_2}} \right) \cr
\uP_{\rm MD} & = 
\cQ \left( \frac{u_1 -\tau}{\sqrt{\sigma_1^2}} \right)
=\cQ \left( \frac{\sqrt{2M} P_{\rm rx} -\tau}{\sqrt{
\rho_1 \rho_2  + P_{\rm rx}^2}}
\right).
	\label{EQ:PPx1}
\end{align}
For convenience,
we can decide $\tau$ so that $\uP_{\rm FA} = 
\uP_{\rm MD}$. In this case,
after some manipulations,
we have
\be
\bar \tau = \frac{N_0 \sqrt{2 M} \gamma}{1 
+ \sqrt{1 + \frac{\gamma^2}{ 
(1 + \kappa_{q_1} \gamma) (1 + \kappa_{q_2} \gamma) }}},
	\label{EQ:btau}
\ee
which leads to
\be
\uP_{\rm FA} = \uP_{\rm MD} = 
\cQ \left(
\sqrt{2M} \theta( \gamma;\kappa_{q_1}, \kappa_{q_2})
 \right),
	\label{EQ:PPx2}
\ee
where
\be
\theta(\gamma,\kappa_{q_1}, \kappa_{q_2}) = \frac{\gamma}{
\kappa(\gamma) + \sqrt{\gamma^2 + \kappa(\gamma)^2}}.
	\label{EQ:theta}
\ee
Here, 
$\kappa (\gamma)= \sqrt{(1+\kappa_{q_1} \gamma)(1+\kappa_{q_1} \gamma)}$.
Since
$\theta(\gamma)$
increases with $\gamma$ and 
it approaches $\frac{1}{
\sqrt{\kappa_{q_1} \kappa_{q_2}} +
\sqrt{1 + \kappa_{q_1} \kappa_{q_2}}}$
as $\gamma \to \infty$, 
$\uP_{\rm FA} = \uP_{\rm MD}$ is lower-bounded 
by $\cQ(\sqrt{2M} c)$, where $c$ is a constant that
is independent of $\gamma$ and $M$.
This demonstrates that 
$M$ has to be large for a low
probability of error.

Since $\{ \kappa_{q_1}, \kappa_{q_2} \}$
are assumed to be known,
there is no decision error when $\min \{ \kappa_{q_1}, \kappa_{q_2} \}
= 0$ (i.e., in this case, without
any decision error, S-preamble $q$ is decided as an unused one).
Thus, the actual probability of FA 
can be lower than that in 
\eqref{EQ:PPx1} or \eqref{EQ:PPx2}.
Furthermore, we note that
the error probability
in \eqref{EQ:PPx2} is a conditional
probability as it depends on
$\{\kappa_{q_1}, \kappa_{q_2} \}$.
Thus, to find the average error probability,
we need the distribution of 
$\{\kappa_{q_1}, \kappa_{q_2} \}$,
which is difficult to derive as 
$\kappa_{q_1}$ and $\kappa_{q_2}$ are not independent.
However, for tractable analysis, we can assume
that they are independent.  Under this assumption,
for convenience, 
we use $\kappa$ that is the sum
of randomly chosen $K$ elements 
of a row of $\bPhi$.
As mentioned earlier, the number of 1's in a row of $\bPhi$
is $L-1$.
Thus,
$\kappa$ can be approximated by a binomial 
random variable as follows:
\be
\kappa \sim {\rm Bin} \left(
\bar K, \frac{L-1}{Q} \right), \ \bar K \le \min\{K, L-1\}.
\ee
Note that the maximum of $\kappa$ is $L-1$. 
Let
$p_k = \binom{\bar K}{k} 
\left(\frac{L-1}{Q} \right)^k \left(1 - \frac{L-1}{Q} \right)^{\bar K-k}$
and
\be
[\bQ]_{k,k^\prime} = 
\cQ \left( \sqrt{2M} \theta( \gamma;k,k^\prime) \right),
\ k,k^\prime \in \{0, \ldots, \bar K\}.
\ee
Then, the average error probability can be approximately found as
\begin{align}
\uE \left[\cQ \left( \sqrt{2M} \theta( \gamma;
\kappa_{q_1},\kappa_{q,2}) \right) \right] 
\approx \bar \uP = \bp^\rT \bQ \bp,
	\label{EQ:bP}
\end{align}
where $\bp = [p_0 \ \ldots \ p_{\bar K}]^\rT$.

\section{Simulation Results}	\label{S:Sim}

In this section, for simulations, it is assumed that 
the channel vectors are generated
according to Assumption {\bf A}).
We present simulation results with finite $M$ and $K$ 
and discuss their impact on the performance.
For comparison, the conventional approach with 
(orthogonal) preambles is considered,
while the approach in \cite{Jiang19} is not 
discussed as its length of preamble transmission phase
is a multiple of $L$, which results in a lower spectral 
efficiency\footnote{For a fair comparison with the approach
in \cite{Jiang19} in terms of throughput as well as spectral efficiency,
the length of data packet has to be fixed.
However, since the length of the data packet can be arbitrary,
comparisons require more careful settings with details,
which will be considered in the future as a further issue.}
than those of the proposed approach
(i.e., $B = 2$) and conventional approach (i.e., $B = 1$).

\begin{figure}[thb]
\begin{center}
\includegraphics[width=\figwidth]{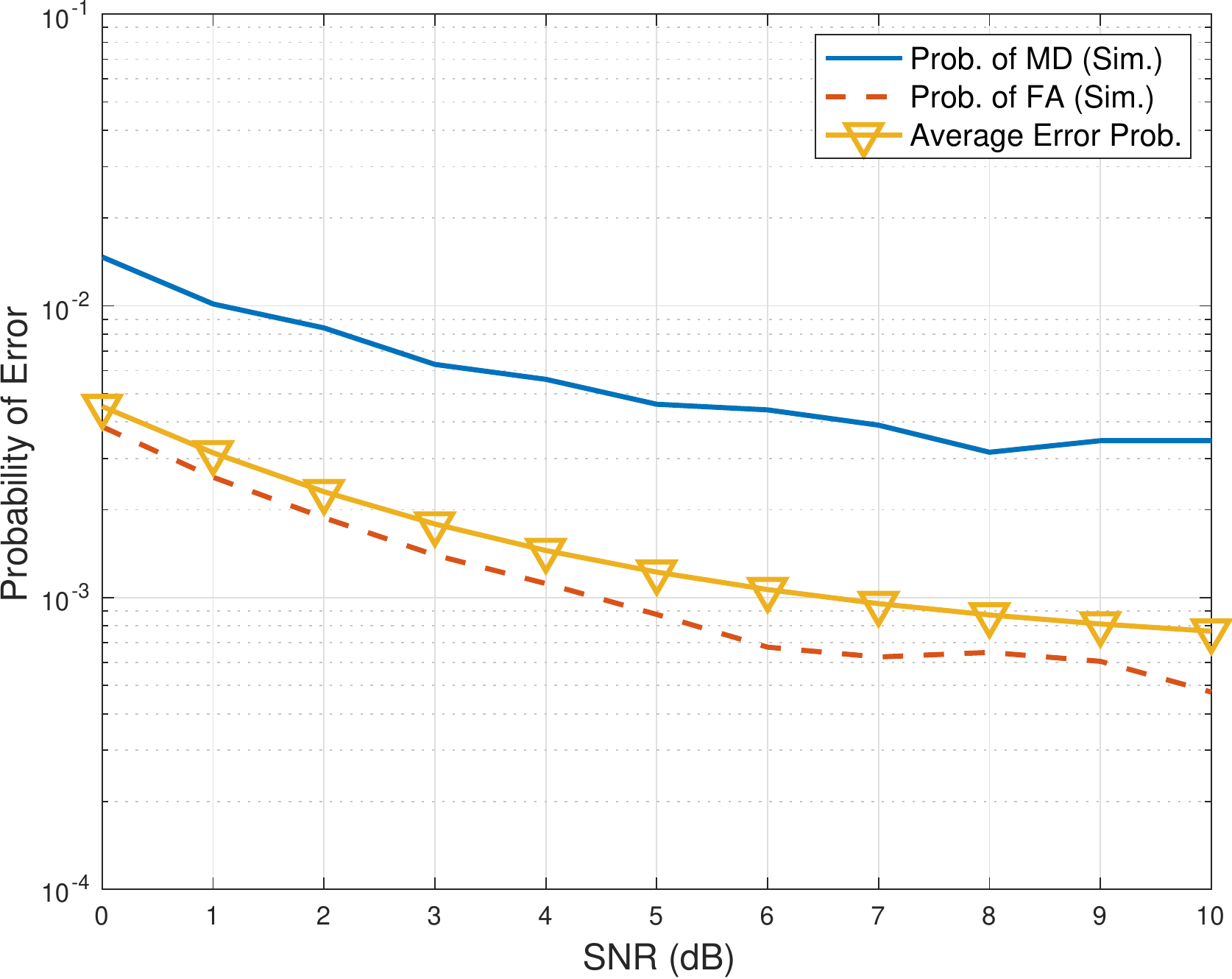}
\end{center}
\caption{Probabilities of MD and FA for S-preamble detection
as functions of SNR, $\gamma$, when $M = 100$, $L = 32$, and $K = 20$.}
        \label{Fig:plt_SPD1}
\end{figure}

We first consider the probabilities of MD and FA
in detecting used or transmitted S-preambles in 
Section~\ref{S:Hyp}.
The decision threshold is set according to \eqref{EQ:btau}.
Fig.~\ref{Fig:plt_SPD1}
shows the error probabilities
as functions of SNR, $\gamma$, when $M = 100$, $L = 32$, and $K = 20$.
In addition to the probabilities of MD and FA obtained from 
simulations, we 
show the average error probability
in \eqref{EQ:bP},
which is an approximate average probability of error
obtained by taking the average with respect to
$\kappa_{q_b}$. As $\gamma$ increases,
it is shown that the probabilities of FA and MD decrease
as expected from \eqref{EQ:PPx2} and \eqref{EQ:theta}. 
Furthermore, as mentioned earlier, it is observed
that the probability of FA is lower than that of MD,
while \eqref{EQ:bP} is shown to be a reasonable approximate.

\begin{figure}[thb]
\begin{center}
\includegraphics[width=\figwidth]{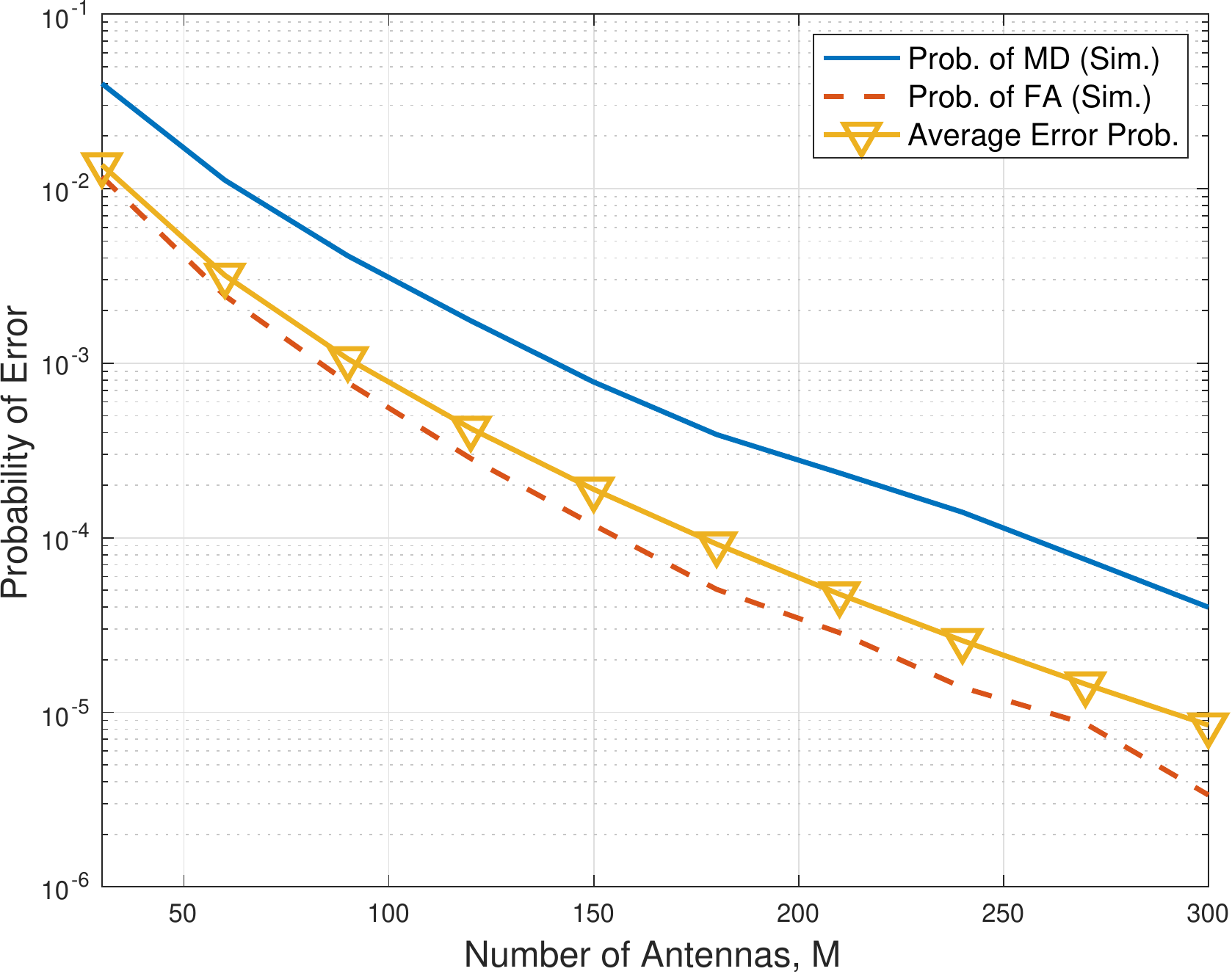}
\end{center}
\caption{Probabilities of MD and FA for S-preamble detection
as functions of $M$ when $K = 20$, $L = 32$, and $\gamma = 10$ dB.}
        \label{Fig:plt_SPD3}
\end{figure}

We present
the probabilities of MD and FA for transmitted S-preamble detection
as functions of $M$ when $K = 20$, $L = 32$, and $\gamma = 10$ dB
in Fig.~\ref{Fig:plt_SPD3}.
As $M$ increases, the probabilities of error decrease,
which can be expected from \eqref{EQ:PPx2} and \eqref{EQ:theta}.

\begin{figure}[thb]
\begin{center}
\includegraphics[width=\figwidth]{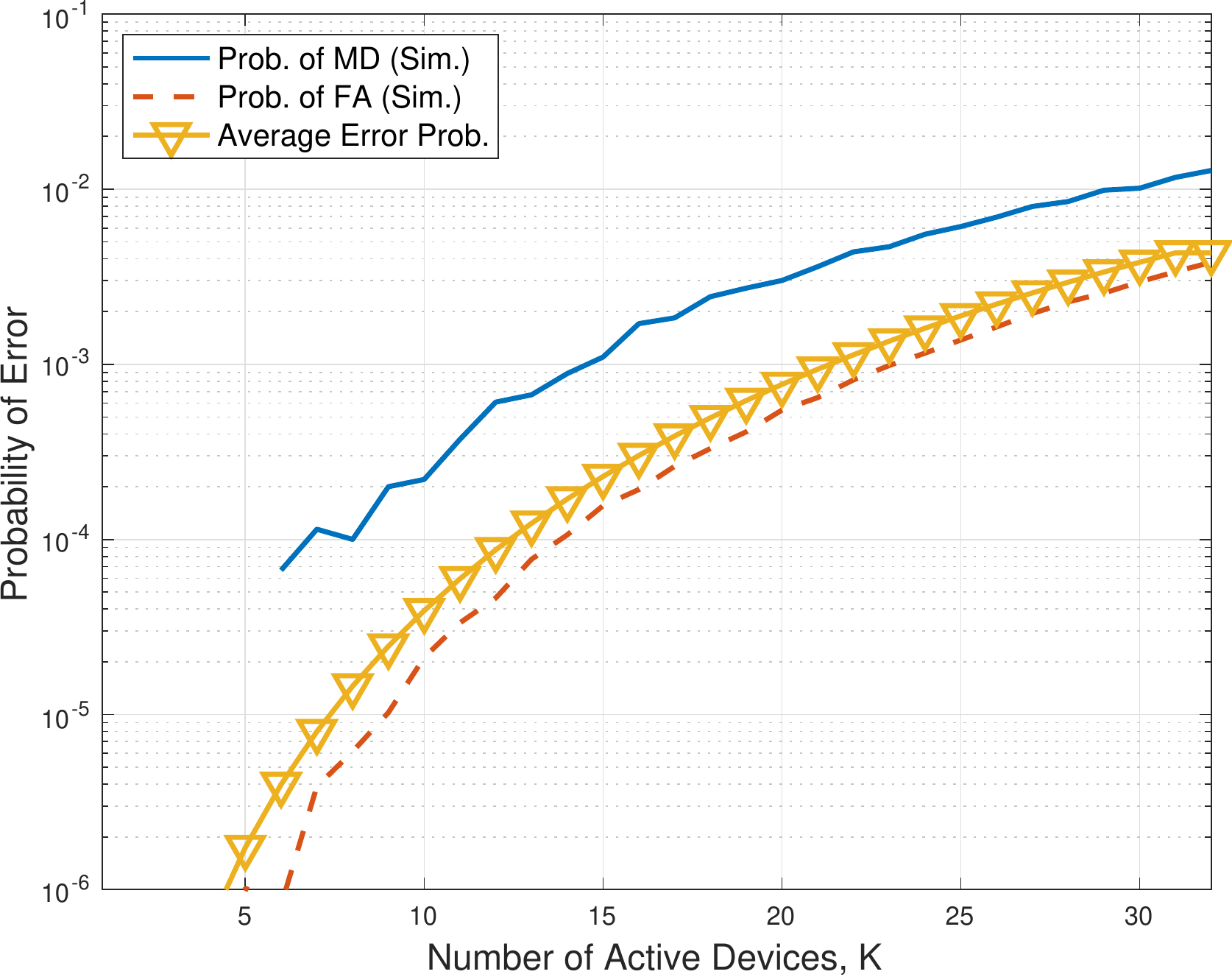}
\end{center}
\caption{Probabilities of MD and FA for S-preamble detection
as functions of $K$ when $M = 100$, $L = 32$, and $\gamma = 10$ dB.}
        \label{Fig:plt_SPD2}
\end{figure}

Fig.~\ref{Fig:plt_SPD2} demonstrates the impact 
of $K$ on the error probabilities
when $M = 100$, $L = 32$, and $\gamma = 10$ dB.
It is observed that as long as $K$ is less than $L$,
the probabilities of MD and FA are sufficiently low.
In other words, the detection of used S-preambles is successful for
the channel estimation with a high probability if $K \le L$,
and the overall performance mainly depends
on the following events: \emph{i)} 
successful decoding (i.e., the SINR is higher than or
equal to the threshold); \emph{ii)} collision of S-preambles.

\begin{figure}[thb]
\begin{center}
\includegraphics[width=\figwidth]{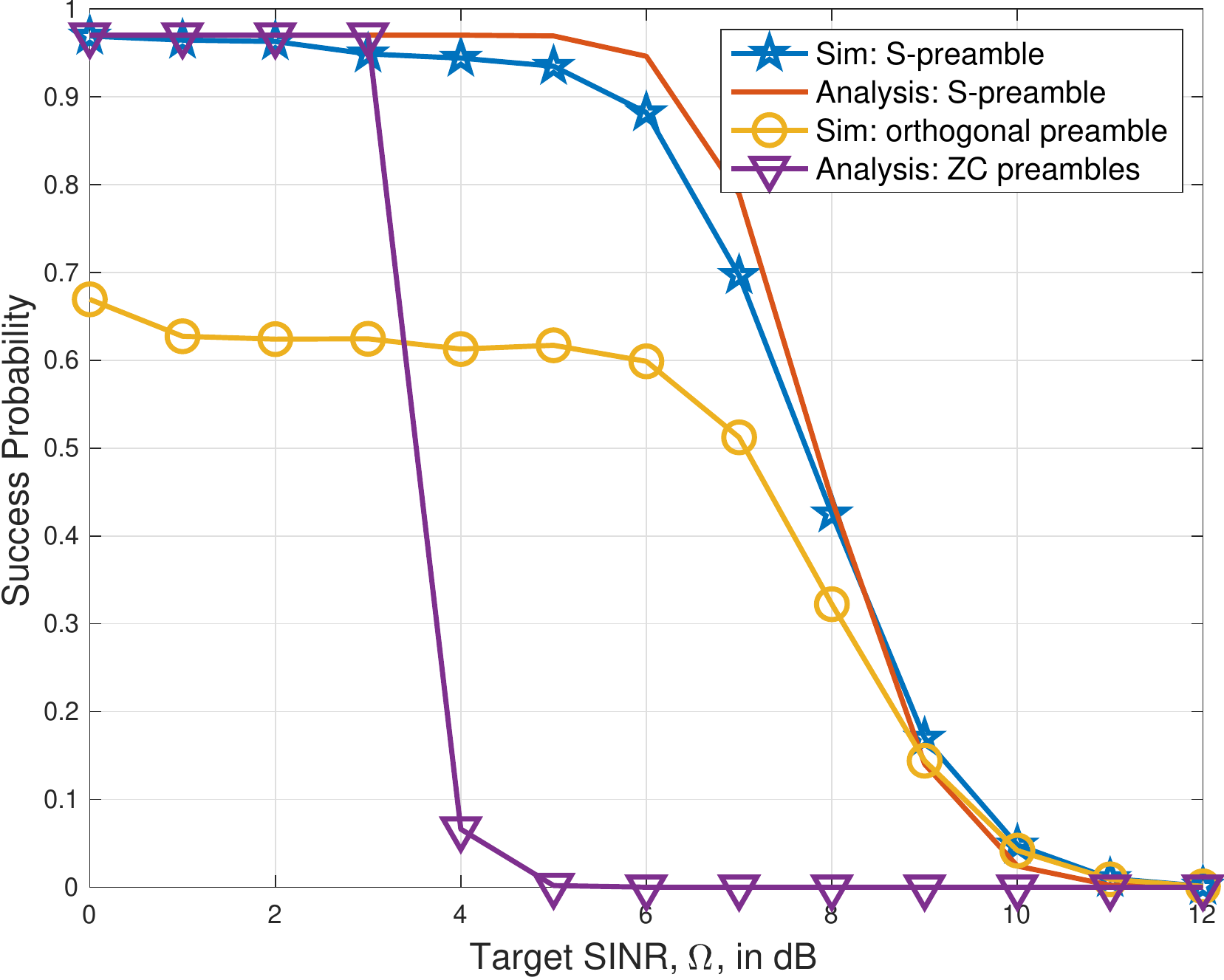}
\end{center}
\caption{Success probability as a function
of the target SINR, $\Omega$,
when $M = 100$, $K = 16$, $L = 32$, and $\gamma = 10$ dB.}
        \label{Fig:plt0}
\end{figure}

\begin{figure}[thb]
\begin{center}
\includegraphics[width=\figwidth]{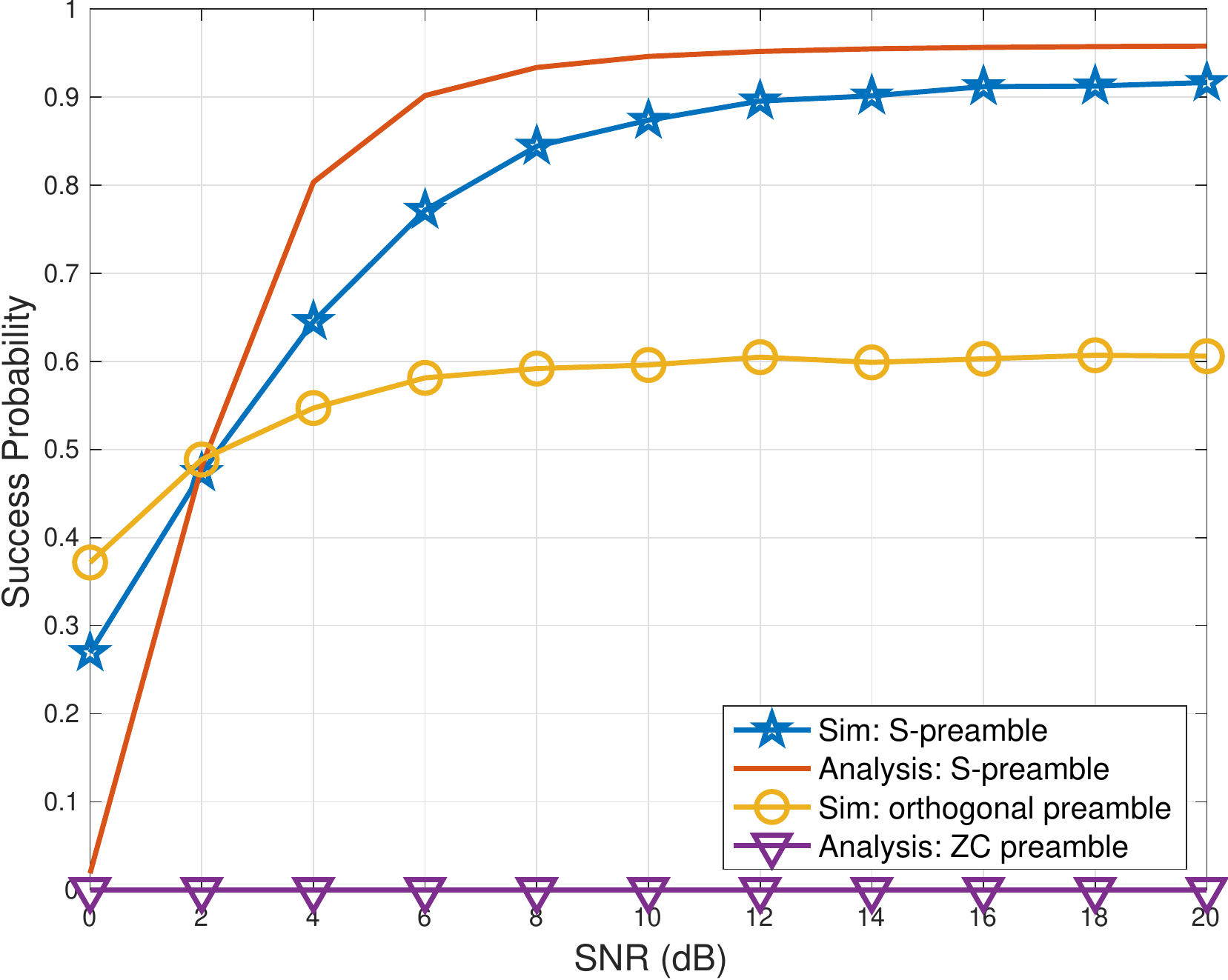}
\end{center}
\caption{Success probability as a function of the SNR, $\gamma$,
when $M = 100$, $K = 16$, $L = 32$, and $\Omega = 6$ dB.}
        \label{Fig:plt1}
\end{figure}

\begin{figure}[thb]
\begin{center}
\includegraphics[width=\figwidth]{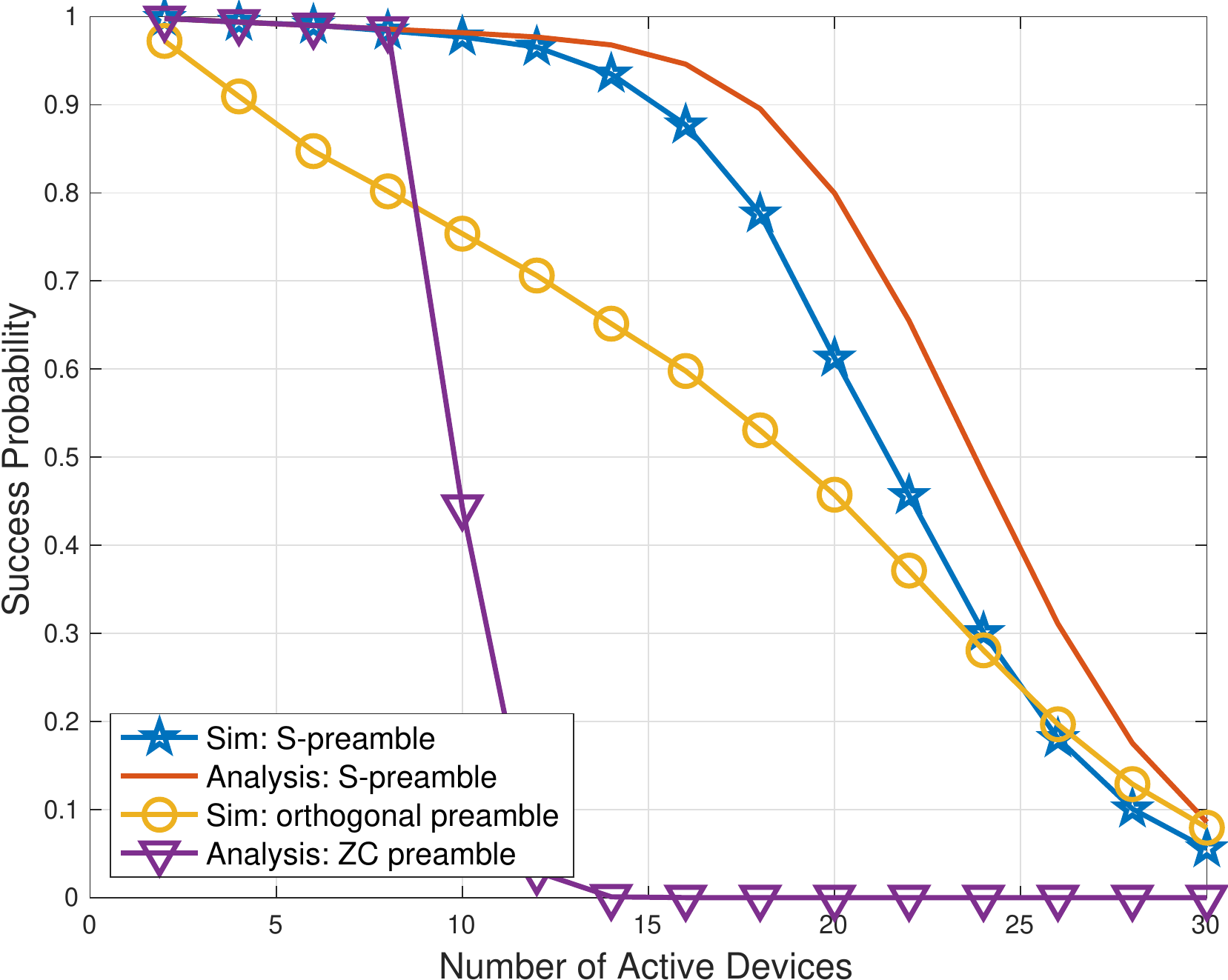}
\end{center}
\caption{Success probability as a function of 
the number of active devices, $K$, 
when $M = 100$, $L = 32$, $\gamma = 10$ dB, and $\Omega = 6$ dB.}
        \label{Fig:plt2}
\end{figure}

\begin{figure}[thb]
\begin{center}
\includegraphics[width=\figwidth]{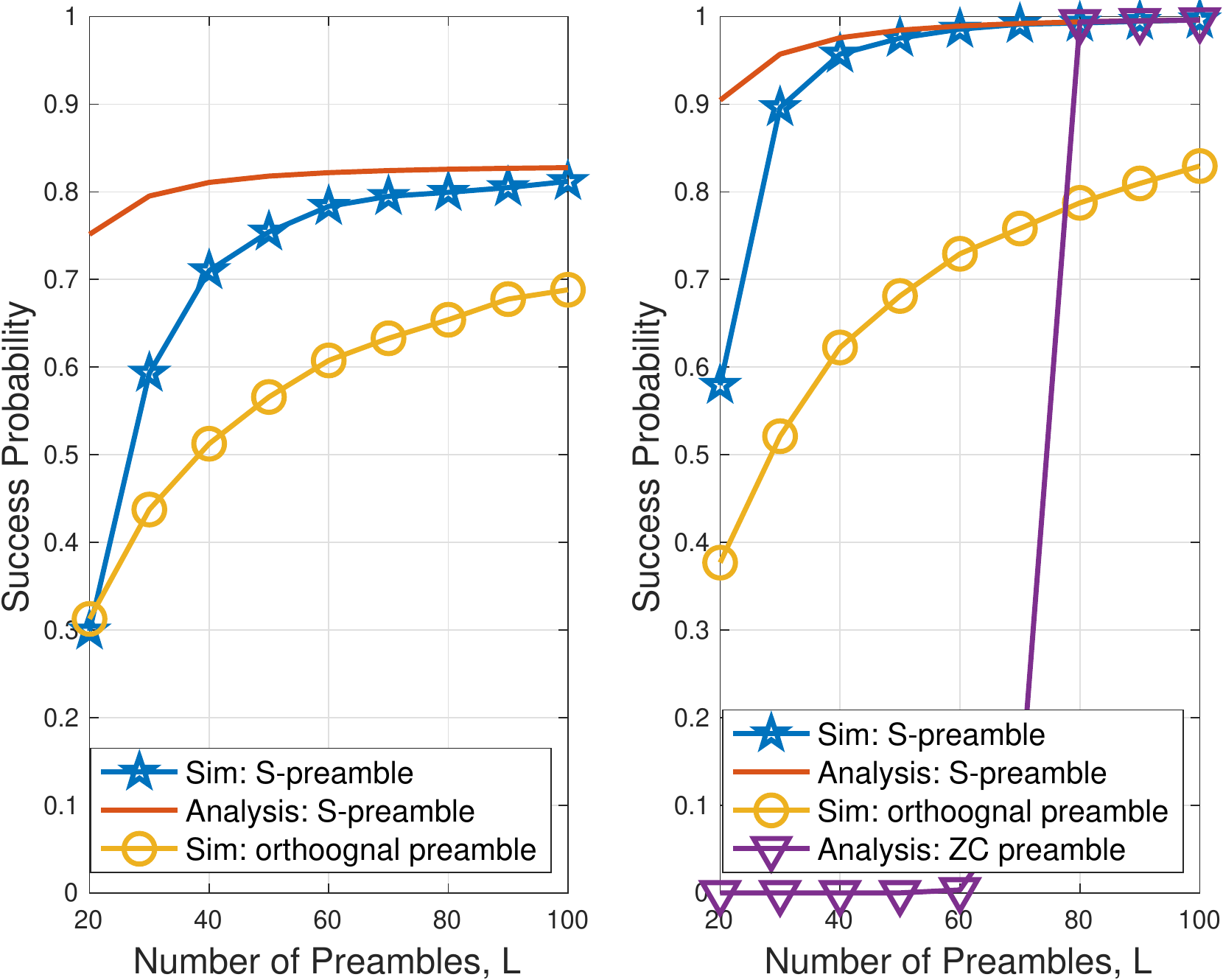} \\
\hskip 0.5cm (a) \hskip 3.5cm (b)
\end{center}
\caption{Success probability as a function of 
the number of preambles, $L$, 
when $K = 20$ and $\Omega = 6$ dB: 
(a) $M = 100$ and $\gamma = 10$ dB; 
(b) $M = 500$ and $\gamma = 30$ dB.}
        \label{Fig:plt3}
\end{figure}

\begin{figure}[thb]
\begin{center}
\includegraphics[width=\figwidth]{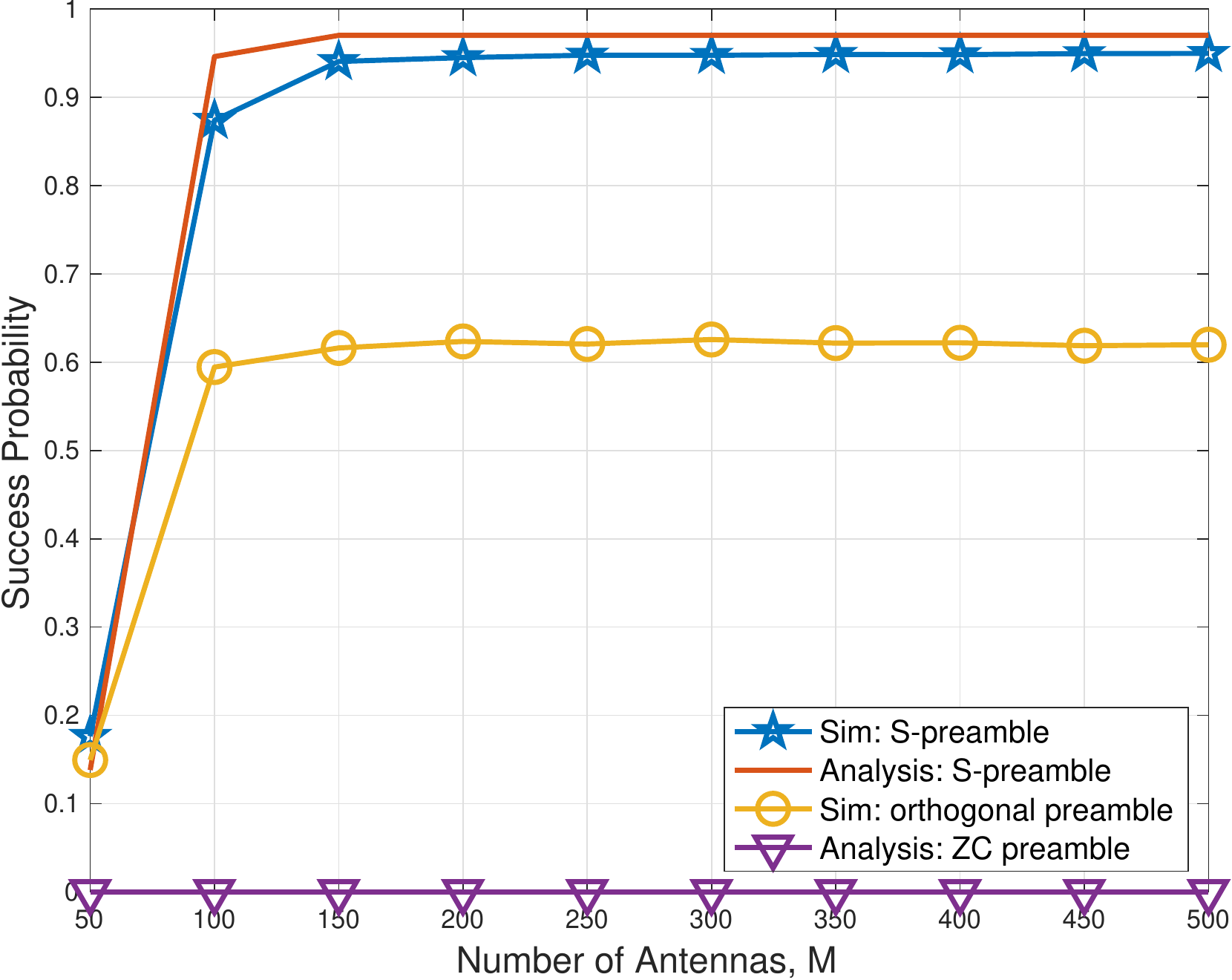}
\end{center}
\caption{Success probability as a function of 
the number of antennas, $M$,
when $K = 16$, $L = 32$, $\gamma = 10$ dB, and $\Omega = 6$ dB.}
        \label{Fig:plt4}
\end{figure}

We now present simulation results for the success
probability, $\eta_k$. Conjugate beamforming
is used with the estimated channel vectors
in both approaches: the conventional approach with 
(orthogonal) preambles (i.e., $B = 1$)
and the proposed approach with S-preambles (i.e., $B = 2$).
In the conventional approach,
since each active device transmits
one preamble, its transmit power becomes doubled
for a fair comparison with the proposed one
where each active device transmits
a superposition of two different preambles, i.e., S-preamble.
The theoretical success probability in \eqref{EQ:eta_k}
is shown with the asymptotic distribution of SINR
obtained using the results in \cite{Ding19_IoT},
which is shown by solid lines in Figs.~\ref{Fig:plt0} --~\ref{Fig:plt4}.
Since this theoretical success probability 
is obtained without taking into consideration 
the rank deficiency shown in Fig.~\ref{Fig:t_mat},
it might be seen as an upper-bound.

In addition, to compare
with the case of conventional
non-orthogonal preambles, we consider $Q$ ZC sequences
of length $L$ and use 
the success probability 
derived in \cite[Eq. 19]{Ding20b},
which is given by
\be
\uP_{\rm succ, ZC} 
= \left(1 - \frac{1}{Q} \right)^{K-1}
\sum_{k=0}^T \binom{K-1}{k} \alpha^k (1- \alpha)^{K-1-k},
	\label{EQ:PZC}
\ee
where $\alpha = 1 - \frac{L}{Q}$
and $T = \min \{\lfloor \frac{L}{\Omega} \rfloor ,K-1\}$.
Actually, \eqref{EQ:PZC} is an asymptotic
probability with $M \to \infty$, in which case
the noise is ignored, i.e., the SINR becomes
the signal-interference ratio (SIR).
As a result, \eqref{EQ:PZC} is independent of $M$
and $N_0$ (or SNR) and can be seen as an upper-bound.

Fig.~\ref{Fig:plt0}
shows the success probability as a function
of the target SINR, $\Omega$,
when $M = 100$, $K = 16$, $L = 32$, and $\gamma = 10$ dB.
In general, the success probability decreases
with $\Omega$, 
while the success probability
of the proposed approach with S-preambles
is higher than that of the conventional approach with preambles
(especially, when $\Omega$ is sufficiently low, i.e.,
$\le 6$ dB). As discussed earlier,
the performance gain results from S-preambles
that can effectively decrease the probability of preamble collision.
It is also shown that
the performance of non-orthogonal ZC preambles
can be significantly degraded 
for $\Omega \ge 4$ dB due to non-zero cross-correlation between
non-orthogonal preambles.
Note that although the length of ZC sequences has to be a prime,
\eqref{EQ:PZC} is valid for any $L$. Thus, \eqref{EQ:PZC} is used for 
the results of ZC preambles in Figs.~\ref{Fig:plt0}--~\ref{Fig:plt4}.

In Fig.~\ref{Fig:plt1}, we show
the success probability as a function of the SNR, $\gamma$,
when $M = 100$, $K = 16$, $L = 32$, and $\Omega = 6$ dB.
As the SNR increases, the success probability
increases. It is noteworthy that
the success probability of the proposed
approach with S-preambles 
is lower than that of the conventional approach
with preambles when $\gamma$ is low (e.g., $\le 2$ dB).
However, since a high success probability
is desirable, it is expected to have a high SNR (e.g., $\ge 10$
dB) in general, where the proposed approach
outperforms the conventional approach.
It is also shown that
the performance of non-orthogonal ZC preambles
is not satisfactory due to a high target SINR,
$\Omega = 6$ dB.

Fig.~\ref{Fig:plt2}
shows
the success probability as a function of 
the number of active devices, $K$, 
when $M = 100$, $L = 32$, $\gamma = 10$ dB, and $\Omega = 6$ dB,
where we can see that the success probability decreases with $K$.
As expected, 
it is also shown that the proposed approach 
can support more active devices than the 
conventional approach. For example, with a success probability
of 0.8, the proposed approach can support around $K = 17$
active devices, while the conventional approach can support around $K = 8$.
The performance of non-orthogonal ZC preambles
is degraded quickly as $K$ increases and becomes
poor when $K \ge 10$.

In Fig.~\ref{Fig:plt3},
the success probability is shown as a function of 
the number of preambles, $L$, 
when $K = 20$, $\Omega = 6$ dB,
and $(M,\gamma) = \{(100, 10 {\rm dB}), 
(500,  30 {\rm dB})\}$. 
As mentioned earlier, the theoretical success
probability does not take into account the rank 
deficiency of $\bar \bPhi$ in Fig.~\ref{Fig:t_mat},
which is more serious when $L$ is close to $K$.
Thus, the theoretical success
probability becomes not tight when $L$ is small. However,
it becomes tight as $L$ increases 
as shown in Fig.~\ref{Fig:plt3}.
Note that the performance of non-orthogonal ZC preambles
from \eqref{EQ:PZC} is an asymptotic result
when $M \to \infty$. 
Thus, as $L \to \infty$ with a fixed $K$,
the asymptotic success probability is only dependent
on preamble collision and becomes
$\left(1 - \frac{1}{Q} \right)^{K-1} \to 1$
as shown in Fig.~\ref{Fig:plt3}.
Thus, \eqref{EQ:PZC} has to be regarded as an upper-bound
and can be used to compare with the proposed
approach when $M$ is sufficiently large,
i.e., the results in Fig.~\ref{Fig:plt3} (b).
In Fig.~\ref{Fig:plt3} (b),
it is clearly shown that $L$ has to be sufficiently large 
for a low cross-correlation when ZC sequences are used 
so that the SIR is higher than or equal to 
the threshold SINR. Otherwise, the success probability
is low.

Finally, Fig.~\ref{Fig:plt4}
shows the success probability as a function of 
the number of antennas, $M$,
when $K = 16$, $L = 32$, $\gamma = 10$ dB, and $\Omega = 6$ dB.
It is shown that the success probability
increases with $M$ in both the proposed and conventional approaches. 
However, although $M \to \infty$,
it is shown that the success probability does not approach 1
in Fig.~\ref{Fig:plt4}
due to preamble collision as discussed earlier. 
In other words, while massive MIMO can provide a high SINR,
its performance is limited by preamble collision.
Since the proposed approach with
S-preambles can effectively reduces
the probability of preamble collision,
it can improve the performance
and has a higher success probability than the conventional 
approach. The success probability in \eqref{EQ:PZC}
for the performance of ZC preambles
is almost 0 as the SINR is lower than $\Omega = 6$ dB
due to non-zero cross-correlation between ZC preambles.

\section{Concluding Remarks}	\label{S:Con}

In this paper, we proposed the use of S-preambles for 
a grant-free random access scheme with massive MIMO
to improve the performance (e.g., 
to support more active devices).
Without increasing the size of preamble pool
and the length of preamble transmission phase,
it was shown that the proposed approach can effectively reduce
preamble collisions using S-preambles and 
increase the success probability.

Since S-preambles can be seen as 
structured \emph{non-orthogonal} preambles,
we believe that there would be different approaches
to design non-orthogonal preambles with effective channel estimation
methods, which might be an interesting topic to be studied in the future.

\bibliographystyle{ieeetr}
\bibliography{mtc}

\begin{thebibliography}{10}

\bibitem{Bockelmann16}
C.~Bockelmann, N.~Pratas, H.~Nikopour, K.~Au, T.~Svensson, C.~Stefanovic,
  P.~Popovski, and A.~Dekorsy, ``Massive machine-type communications in {5G}:
  physical and {MAC}-layer solutions,'' {\em IEEE Communications Magazine},
  vol.~54, pp.~59--65, Sep 2016.

\bibitem{Dawy17}
Z.~{Dawy}, W.~{Saad}, A.~{Ghosh}, J.~G. {Andrews}, and E.~{Yaacoub}, ``Toward
  massive machine type cellular communications,'' {\em IEEE Wireless
  Communications}, vol.~24, pp.~120--128, Feb 2017.

\bibitem{3GPP_MTC}
3GPP TR 37.868 V11.0, {\em Study on {RAN} improvments for machine-type
  communications}, October 2011.

\bibitem{3GPP_NBIoT}
3GPP TS 36.321 V13.2.0, {\em Evolved Universal Terrestrial Radio Access
  ({E-UTRA}); Medium Access Control ({MAC}) protocol specification}, June 2016.

\bibitem{Chang15}
C.~H. Chang and R.~Y. Chang, ``Design and analysis of multichannel slotted
  {ALOHA} for machine-to-machine communication,'' in {\em Proc. IEEE GLOBECOM},
  pp.~1--6, Dec 2015.

\bibitem{Choi16}
J.~Choi, ``On the adaptive determination of the number of preambles in {RACH}
  for {MTC},'' {\em IEEE Communications Letters}, vol.~20, pp.~1385--1388, July
  2016.

\bibitem{Marzetta10}
T.~L. Marzetta, ``Noncooperative cellular wireless with unlimited numbers of
  base station antennas,'' {\em IEEE Trans. Wireless Communications}, vol.~9,
  pp.~3590--3600, Nov. 2010.

\bibitem{Bjornson18}
E.~{Björnson}, J.~{Hoydis}, and L.~{Sanguinetti}, ``Massive {MIMO} has
  unlimited capacity,'' {\em IEEE Trans. Wireless Communications}, vol.~17,
  pp.~574--590, Jan 2018.

\bibitem{deC17}
E.~{de Carvalho}, E.~{Björnson}, J.~H. {Sørensen}, E.~G. {Larsson}, and
  P.~{Popovski}, ``Random pilot and data access in massive {MIMO} for
  machine-type communications,'' {\em IEEE Trans. Wireless Communications},
  vol.~16, pp.~7703--7717, Dec 2017.

\bibitem{Senel17}
K.~Senel and E.~G. Larsson, ``Device activity and embedded information bit
  detection using {AMP} in massive {MIMO},'' in {\em 2017 IEEE Globecom
  Workshops (GC Wkshps)}, pp.~1--6, Dec 2017.

\bibitem{Liu18}
L.~Liu, E.~G. Larsson, W.~Yu, P.~Popovski, C.~Stefanovic, and E.~de~Carvalho,
  ``Sparse signal processing for grant-free massive connectivity: A future
  paradigm for random access protocols in the {I}nternet of {T}hings,'' {\em
  IEEE Signal Processing Magazine}, vol.~35, pp.~88--99, Sept 2018.

\bibitem{Ding19_IoT}
J.~{Ding}, D.~{Qu}, H.~{Jiang}, and T.~{Jiang}, ``Success probability of
  grant-free random access with massive {MIMO},'' {\em IEEE Internet of Things
  J.}, vol.~6, pp.~506--516, Feb 2019.

\bibitem{Bockelmann18}
C.~{Bockelmann}, N.~K. {Pratas}, G.~{Wunder}, S.~{Saur}, M.~{Navarro},
  D.~{Gregoratti}, G.~{Vivier}, E.~{De Carvalho}, Y.~{Ji}, C.~{Stefanović},
  P.~{Popovski}, Q.~{Wang}, M.~{Schellmann}, E.~{Kosmatos}, P.~{Demestichas},
  M.~{Raceala-Motoc}, P.~{Jung}, S.~{Stanczak}, and A.~{Dekorsy}, ``Towards
  massive connectivity support for scalable {mMTC} communications in {5G}
  networks,'' {\em IEEE Access}, vol.~6, pp.~28969--28992, 2018.

\bibitem{Choi17IoT}
J.~Choi, ``Two-stage multiple access for many devices of unique identifications
  over frequency-selective fading channels,'' {\em IEEE Internet of Things J.},
  vol.~4, pp.~162--171, Feb 2017.

\bibitem{Choi20b}
J.~{Choi}, ``On throughput of compressive random access for one short message
  delivery in {IoT},'' {\em IEEE Internet of Things J.}, vol.~7, no.~4,
  pp.~3499--3508, 2020.

\bibitem{Jiang19}
H.~{Jiang}, D.~{Qu}, J.~{Ding}, and T.~{Jiang}, ``Multiple preambles for high
  success rate of grant-free random access with massive {MIMO},'' {\em IEEE
  Trans. Wireless Communications}, vol.~18, pp.~4779--4789, Oct 2019.

\bibitem{Ding20b}
J.~{Ding}, D.~{Qu}, and J.~{Choi}, ``Analysis of non-orthogonal sequences for
  grant-free {RA} with massive {MIMO},'' {\em IEEE Trans. Communications},
  vol.~68, pp.~150--160, Jan 2020.

\bibitem{Donoho06}
D.~Donoho, ``Compressed sensing,'' {\em IEEE Trans. Information Theory},
  vol.~52, pp.~1289--1306, April 2006.

\bibitem{Candes06}
E.~Candes, J.~Romberg, and T.~Tao, ``Robust uncertainty principles: exact
  signal reconstruction from highly incomplete frequency information,'' {\em
  IEEE Trans. Information Theory}, vol.~52, pp.~489--509, Feb 2006.

\bibitem{Seo19}
H.~{Seo}, J.~{Hong}, and W.~{Choi}, ``Low latency random access for sporadic
  {MTC} devices in internet of things,'' {\em IEEE Internet of Things J.},
  vol.~6, pp.~5108--5118, June 2019.

\bibitem{Choi19c}
J.~{Choi}, ``{NOMA}-based compressive random access using {G}aussian
  spreading,'' {\em IEEE Trans. Communications}, vol.~67, no.~7,
  pp.~5167--5177, 2019.

\bibitem{Bjornson16}
E.~{Björnson}, E.~G. {Larsson}, and M.~{Debbah}, ``Massive {MIMO} for maximal
  spectral efficiency: How many users and pilots should be allocated?,'' {\em
  IEEE Trans. Wireless Communications}, vol.~15, pp.~1293--1308, Feb 2016.

\bibitem{Asratian98}
A.~S. Asratian, T.~M.~J. Denley, and R.~H\"{a}ggkvist, {\em Bipartite Graphs
  and Their Applications}.
\newblock New York, NY, USA: Cambridge University Press, 1998.

\bibitem{Scharf91}
L.~Scharf and C.~Demeure, {\em Statistical Signal Processing: Detection,
  Estimation, and Time Series Analysis}.
\newblock Addison-Wesley series in electrical and computer engineering,
  Addison-Wesley Publishing Company, 1991.

\bibitem{Kay98}
S.~Kay, {\em Fundamentals of Statistical Signal Processing: Detection theory}.
\newblock Prentice Hall Signal Processing Series, Prentice-Hall PTR, 1998.

\bibitem{Mallik11}
R.~K. {Mallik} and N.~C. {Sagias}, ``Distribution of inner product of complex
  gaussian random vectors and its applications,'' {\em IEEE Trans.
  Communications}, vol.~59, pp.~3353--3362, December 2011.

\bibitem{Picinbono96}
B.~{Picinbono}, ``Second-order complex random vectors and normal
  distributions,'' {\em IEEE Trans. Signal Processing}, vol.~44,
  pp.~2637--2640, Oct 1996.

\bibitem{Bos95}
A.~{van den Bos}, ``The multivariate complex normal distribution-a
  generalization,'' {\em IEEE Trans. Information Theory}, vol.~41,
  pp.~537--539, March 1995.

\bibitem{Duda00}
R.~O. Duda, P.~E. Hart, and D.~G. Stork, {\em Pattern Classification (2nd
  Edition)}.
\newblock New York, NY, USA: Wiley-Interscience, 2000.

\end{thebibliography}

\end{document}